\documentclass[12pt]{article}
\usepackage[margin=2.5cm]{geometry}
\usepackage{amsmath}
\usepackage{amssymb}
\usepackage[utf8]{inputenc}
\usepackage{bbold}
\usepackage{xcolor}
\usepackage[makeroom]{cancel}
\usepackage{mathtools}
\usepackage{bbm}
\usepackage[colorlinks, allcolors=cyan]{hyperref}

\usepackage[backend=biber, style=phys, sorting=none, citestyle=numeric-comp, url=false]{biblatex}
\appto{\bibsetup}{\sloppy}

\bibliography{bibliography}

\title{Unification of Conformal and Fuzzy Gravities with Internal Interactions -- study of their behaviour in low energies and possible signals in the detection of Gravitational Waves.}
\begin{document}
\author{Gregory Patellis$^1$, Danai Roumelioti$^1$, Stelios Stefas$^1$\thanks{Corresponding author.}, George Zoupanos$^{1,2,3,4}$}\date{}

\maketitle
\begin{center}
\itshape$^1$ Physics Department, National Technical University of Athens, Zografou Campus, 157 80, Zografou, Greece\\
\itshape$^2$ Max-Planck Institut f\'ur Physik, Boltzmannstr. 8, 85 748 Garching/Munich, Germany\\
\itshape$^3$ Universit\"at Hamburg, Luruper Chaussee 149, 22 761 Hamburg, Germany\\
\itshape$^4$  Deutsches Elektronen-Synchrotron DESY, Notkestra{\ss}e 85, 22 607, Hamburg, Germany
\end{center}

\begin{center}
\emph{E-mails: \href{mailto:grigorios.patellis@tecnico.ulisboa.pt}{grigorios.patellis@tecnico.ulisboa.pt}, \href{mailto:danai_roumelioti@mail.ntua.gr}{danai\_roumelioti@mail.ntua.gr}, \href{mailto:dstefas@mail.ntua.gr}{dstefas@mail.ntua.gr}, \href{mailto:george.Zoupanos@cern.ch}{george.zoupanos@cern.ch}}
\end{center}

\begin{abstract}
The Unification of Conformal and Fuzzy gravities with Internal Interactions is based on the following two facts. The first is that the tangent group of a curved manifold and the manifold itself do not necessarily have the same dimensions. The second is that both gravitational theories considered here have been formulated in a gauge theoretic way. Here we would like to start by reviewing the gauge theoretic approach of gravities commenting in particular their diffeomorphism invariance. Then we will review the construction of the Conformal Gravity and the Noncommutative (Fuzzy) Gravity using the gauge theoretic framework. Finally based on an extension of the four-dimensional tangent group we will present the Unification of both Gravities with the Internal Interactions. Both unified schemes will be examined further concerning their behaviour in low energies after suitable spontaneous symmetry breakings as well as the possible signals of the related cosmic strings in the gravitational waves.
\end{abstract}

\section{Introduction}
\label{sec1}

The unification of all fundamental interactions has been a central objective of theoretical physics for over a century. One of the earliest attempts was made by Kaluza and Klein~\cite{Kaluza:1921,Klein:1926}, who proposed a framework unifying gravity and electromagnetism---the two well-established interactions at that time---by extending spacetime to five dimensions. Their idea was to reduce a purely gravitational theory defined in five dimensions to four, thereby generating a $U(1)$ gauge theory interpreted as electromagnetism, coupled to gravity. Although initially considered speculative due to its higher-dimensional nature, this approach gained renewed interest when it was realized that non-Abelian gauge theories could naturally emerge from similar settings~\cite{Kerner:1968, CHO1987358, Cho:1975sf} and could be useful in the description of the Standard Model (SM) of particle physics. Specifically, it was found that a higher-dimensional spacetime of the form $M_D = M_4 \times B$, with $B$ a compact Riemannian manifold possessing non-Abelian isometry group $S$, leads upon dimensional reduction to a four-dimensional theory featuring Einstein gravity coupled to a Yang--Mills gauge theory with group $S$, along with scalar fields.

The primary appeal of this construction lies in its geometrical origin of gauge interactions, offering a natural explanation for the appearance of gauge symmetries. However, this framework faces notable obstacles, including the lack of a viable classical ground state with a simple direct product structure and, more crucially for low-energy physics, its inability to yield chiral fermions in four dimensions after dimensional reduction~\cite{Witten:1983}. A remarkable improvement arises when Yang--Mills fields are incorporated from the outset, albeit at the cost of abandoning a purely geometric unification.

In higher-dimensional Grand Unified Theory (GUT) frameworks that include both Yang--Mills fields and fermions~\cite{Georgi:1974sy, FRITZSCH1975193}, the emergence of chiral fermions in four dimensions requires the total spacetime dimension to be of the form $4n + 2$~\cite{CHAPLINE1982461}. In the present work, we primarily explore approaches that move towards the opposite direction -- approaches that treat all interactions, including gravity, as manifestations of gauge symmetries. Nevertheless, Superstring Theories have long dominated the investigation of higher-dimensional unification~\cite{Green2012-ul, polchinski_1998, Lust:1989tj}.

It is fair to say that Superstring Theories (SSTs) offer a consistent framework in higher dimensions, with the heterotic string theory~\cite{GROSS1985253}—formulated in ten dimensions—standing out as a particularly attractive case. This theory naturally accommodates Grand Unified Theory (GUT) gauge groups such as $E_8 \times E_8$, whose dimensional reduction can, in principle, reproduce the SM. It must be emphasized, however, that experimental confirmation of these frameworks is still lacking.

Even prior to the emergence of SSTs, an alternative program was developed based on the dimensional reduction of higher-dimensional gauge theories~\cite{forgacs, MANTON1981502, Kubyshin:1989vd, KAPETANAKIS19924, LUST1985309}. While less ambitious in scope—since it effectively neglects gravity as a dynamical interaction—this approach shared the overarching goal of unifying the remaining fundamental forces. Within this context, Forgacs and Manton (F-M) introduced the Coset Space Dimensional Reduction (CSDR) scheme~\cite{forgacs, MANTON1981502, Kubyshin:1989vd, KAPETANAKIS19924}, wherein chiral fermions can emerge naturally. In parallel, Scherk and Schwarz (S-S) developed the group manifold reduction approach~\cite{SCHERK197961}, which, despite the fact that it cannot accommodate chiral fermions, essential for viable low-energy models, it nevertheless inspired numerous developments in string model building.

More recently, substantial efforts have been dedicated to constructing phenomenologically viable models within the CSDR framework, which from the outset appeared far more promising~\cite{Manousselis_2004, Chatzistavrakidis:2009mh, Irges:2011de, Manolakos:2020cco, Patellis:2024dfl}.

An important line of research that aligns closely with the framework that will be presented in this paper emerged within four-dimensional spacetime and builds on the natural link between gravity and gauge theories. The SM of particle physics, which is a highly successful gauge theory verified extensively in past and current experiments (notably at the LHC), exemplifies this paradigm. On the other hand, it has long been known that gravity too can be formulated as a gauge theory~\cite{utiyama, kibble1961, Sciama, Umezawa, Matsumoto, macdowell, Ivanov:1980tw, Ivanov:1981wn, stellewest, Kibble:1985sn}. Interest in this formulation was reignited by developments in supergravity~\cite{freedman_vanproeyen_2012, Ortín_2015}, which also rely on gauge principles. These ideas have since been extended to Noncommutative (NC) gravity~\cite{castellani, Chatzistavrakidis_2018, Manolakos_paper1, manolakosphd, Manolakos_paper2, Manolakos:2022universe, Manolakos:2023hif, roumelioti2407}.

Weyl~\cite{weyl, weyl1929} was the first to relate electromagnetism to local phase transformations of the electron field and introduced the vierbein formalism, which later became crucial in the gauge formulation of gravity. Utiyama~\cite{utiyama} made the next, decisive step, by showing that gravity could be regarded as a gauge theory of the Lorentz group $SO(1,3)$, though his introduction of the vierbein was somewhat ad hoc. This `weakness' was resolved by Kibble~\cite{kibble1961} and Sciama~\cite{Sciama}, who advocated gauging the full Poincar\'e group. Further advances by Stelle and West~\cite{stellewest, Kibble:1985sn} led to more elegant constructions based on the de Sitter ($SO(1,4)$) or anti-de Sitter ($SO(2,3)$) groups, incorporating spontaneous symmetry breaking (SSB) to recover Lorentz invariance. The conformal group $SO(2,4)$ also played a pivotal role in the formulation of Weyl Gravity (WG)~\cite{KAKU1977304, Roumelioti:2024lvn}, Fuzzy Gravity (FG)~\cite{Chatzistavrakidis_2018, Manolakos_paper1, manolakosphd, Manolakos_paper2, Manolakos:2022universe, Manolakos:2023hif, roumelioti2407}, and their supersymmetric extensions in $\mathcal{N} = 1$ supergravity~\cite{KAKU1977304, freedman_vanproeyen_2012}.

A more direct and ambitious unification strategy involves embedding gravity within a larger gauge group that also includes internal symmetries of particle physics on equal footing~\cite{Weinberg:1984ke, Percacci:1984ai, Percacci_1991}. Recent (and current) work has revitalized this direction~\cite{Nesti_2008, Nesti_2010, Chamseddine2010, Chamseddine2016, Krasnov:2017epi, Konitopoulos:2023wst, Manolakos:2023hif, noncomtomos, Roumelioti:2024lvn, Patellis:2024znm, Roumelioti:dubna, Roumelioti:2025cxi,Patellis:2025qbl}, capitalizing on the observation that the tangent group of a curved spacetime need not match the manifold's dimensionality. This opens the door to using higher-dimensional tangent groups in four-dimensional spacetime, facilitating a unified gauge-theoretic treatment of gravity and internal interactions. In this context, methods developed for higher-dimensional theories, such as CSDR~\cite{CHAPLINE1982461, forgacs, MANTON1981502, Kubyshin:1989vd, KAPETANAKIS19924, LUST1985309, SCHERK197961, Manousselis_2004, Chatzistavrakidis:2009mh, Irges:2011de, Manolakos:2020cco, Patellis:2024dfl}, can be adapted to these four-dimensional constructions. Furthermore, challenges such as implementing simultaneous Weyl and Majorana conditions to obtain realistic chiral spectra also reappear in this setting~\cite{CHAPLINE1982461, KAPETANAKIS19924}. Recently, a unified gauge framework has been constructed combining conformal gravity and internal interactions~\cite{Manolakos:2023hif, Konitopoulos:2023wst, Patellis:2024znm, Roumelioti:2024lvn, Roumelioti:dubna, Roumelioti:2025cxi, Patellis:2025qbl} and extending to noncommutative (fuzzy) spaces~\cite{roumelioti2407}.

In the present work, we begin by presenting the gauge-theoretic formulation of gravity. We then present the construction of Conformal Gravity and NC (Fuzzy) Gravity within this framework. Finally, we develop a unified description of conformal and fuzzy gravity with $SO(10)$ internal symmetries, based on the larger gauge group $SO(2,16)$. It is worth recalling that the spinor representation of $SO(10)$ naturally incorporates right-handed neutrinos, thus allowing the introduction of neutrino masses. We conclude with an estimation of the  symmetry-breaking channels of $SO(10)$, evaluating their compatibility with proton lifetime bounds and their potential observability via gravitational signals from cosmic string production.

\section{Conformal Gauge Gravity}
As previously mentioned, Einstein Gravity (EG) has commonly been viewed through the lens of a gauge theory constructed from the Poincaré group. Nonetheless, a deeper and more precise formulation arises when one instead examines gauge theories associated with the de Sitter (dS), $SO(1,4)$, and anti-de Sitter (AdS), $SO(2,3)$, groups. Similarly to the Poincaré group, these groups also contain 10 generators and can undergo spontaneous symmetry breaking to the Lorentz group, $SO(1,3)$, via non-dynamical (auxiliary) scalar fields \cite{stellewest, Kibble:1985sn, Roumelioti:2024lvn, manolakosphd}. The dS and AdS groups are subgroups of the larger conformal group $SO(2,4)$, which has 15 generators and maintains the invariance of the null interval $ds^2 = \eta_{\mu\nu} dx^\mu dx^\nu = 0$ under transformations in spacetime. In \cite{Kaku:1978nz}, the gravitational gauge theory paradigm was extended to incorporate this full conformal group, giving rise to what is known as Conformal Gravity (CG). Traditionally, transitions from CG to either EG or Weyl-invariant gravity have been implemented through constraint conditions (as seen in \cite{Kaku:1978nz}). By contrast, the method introduced in \cite{Roumelioti:2024lvn} offers a new perspective by achieving spontaneous symmetry breaking of the conformal gauge group. This is realized by including a scalar field within the action, which attains a non-zero vacuum expectation value (vev) via the application of a Lagrange multiplier.

\subsection{Spontaneous symmetry breaking by introducing a scalar in the adjoint representation}
\label{sec2.1}

The conformal gauge group $SO(2,4)$ comprises fifteen generators. Within a four-dimensional setting, these are categorized into six Lorentz generators $M_{ab}$, four generators associated with spacetime translations $P_a$, four special conformal (or conformal boost) generators $K_a$, and one generator corresponding to dilatations $D$.

The gauge connection $A_\mu$, which resides in the Lie algebra of $SO(2,4)$, is constructed as a linear combination of these generators:
\begin{equation}
A_\mu= \frac{1}{2}\omega_\mu{}^{a b} M_{a b}+e_\mu{}^a P_a+b_\mu{}^a K_a+\tilde{a}_\mu D,
\end{equation}
where each component gauge field is associated with a specific generator. In particular, $e_\mu{}^a$ is interpreted as the vierbein (or tetrad), and $\omega_\mu{}^{ab}$ serves as the spin connection. The field strength tensor corresponding to $A_\mu$ is then defined as:
\begin{equation}\label{fst}
F_{\mu \nu}=\frac{1}{2}R_{\mu \nu}{}^{a b} M_{a b}+\tilde{R}_{\mu \nu}{}^a P_a+R_{\mu \nu}{}^a K_a+R_{\mu \nu} D,
\end{equation}
where
\begin{equation}\label{curves}
\begin{aligned}
R_{\mu \nu}{}^{a b} & =\partial_\mu \omega_\nu{}^{a b}-\partial_\nu \omega_\mu{}^{a b}-\omega_\mu{}^{a c} \omega_{\nu c}{}^b+\omega_\nu{}^{a c} \omega_{\mu c}{}^b-8 e_{[\mu}{}^{[a} b_{\nu]}{}^{b]} \\
& =R_{\mu \nu}^{(0) a b}-8 e_{[\mu}{}^{[a} b_{\nu]}{}^{b]}, \\
\tilde{R}_{\mu \nu}{}^a & =\partial_\mu e_\nu{}^a-\partial_\nu e_\mu{}^a+\omega_\mu{}^{a b} e_{\nu b}-\omega_\nu{}^{a b} e_{\mu b}-2 \tilde{a}_{[\mu} e_{\nu]}{}^a \\
& =T_{\mu \nu}^{(0) a}(e)-2 \tilde{a}_{[\mu} e_{\nu]}{}^a, \\
R_{\mu \nu}{}^a & =\partial_\mu b_\nu{}^a-\partial_\nu b_\mu{}^a+\omega_\mu{}^{a b} b_{\nu b}-\omega_\nu{}^{a b} b_{\mu b}+2 \tilde{a}_{[\mu} b_{\nu]}{}^a\\
&=T_{\mu \nu}^{(0) a}(b)+2 \tilde{a}_{[\mu} b_{\nu]}{}^a,\\
R_{\mu \nu} & =\partial_\mu \tilde{a}_\nu-\partial_\nu \tilde{a}_\mu+4 e_{[\mu}{}^a b_{\nu] a},
\end{aligned}
\end{equation}
Here, $T_{\mu \nu}^{(0) a}(e)$ and $R_{\mu \nu}^{(0) a b}$ represent the torsion and curvature in the standard vierbein formalism of General Relativity (GR), while $T_{\mu \nu}^{(0) a}(b)$ corresponds to the torsion related to the auxiliary gauge field $b_\mu{}^a$.

To proceed, we adopt a parity-preserving action that is quadratic in the field strength tensor \eqref{fst}. We also introduce an auxiliary scalar field transforming in the adjoint (15-dimensional) representation of $SO(6) \sim SO(2,4)$, and include a mass parameter $m$:
\begin{equation}
S_{SO(2,4)}=a_{CG}\int d^4x [\operatorname{tr} \epsilon^{\mu \nu \rho \sigma} m\phi F_{\mu \nu}F_{\rho \sigma}+\lambda(\phi^2-m^{-2} \mathbbm{1}_4) ],
\end{equation}
where the trace is over the generators in accordance with their gauge representation.

This scalar auxiliary field $\phi$, expanded on the generators of the gauge algebra, can be written as:
\begin{equation}
\phi=\phi^{a b} M_{a b}+\tilde{\phi}^a P_a+\phi^a K_a+\tilde{\phi} D,
\end{equation}

Following the strategy used in \cite{Li:1973mq}, we fix the gauge so that $\phi$ becomes diagonal and aligned solely along the dilatation generator $D$, specifically taking the form $\operatorname{diag}(1,1,-1,-1)$. Under the constraint $\phi^2 = m^{-2} \mathbbm{1}_4$, this reduces to:
\begin{equation}
\phi=\phi^0=\tilde{\phi}D \xrightarrow{\phi^2=m^{-2}\mathbbm{1}_4}\phi=-2m^{-1} D.
\end{equation}
Substituting this gauge choice into the action yields:
\begin{equation}
S=-2a_{CG}\int d^4x \operatorname{tr} \epsilon^{\mu \nu \rho \sigma} F_{\mu \nu}F_{\rho \sigma}D.
\end{equation}
Because of spontaneous symmetry breaking, the fields $e_\mu{}^a$, $b_\mu{}^a$, and $\tilde{a}_\mu$ must all be rescaled as $me_\mu{}^a$, $mb_\mu{}^a$, and $m\tilde{a}_\mu$ respectively. Substituting the expanded field strength tensor from equation \eqref{fst}, and applying the generator algebra, leads to:
\begin{equation}
\begin{gathered}
S=-2a_{CG}\int d^4x \operatorname{tr} \epsilon^{\mu \nu \rho \sigma}\Big[\frac{1}{4}R_{\mu \nu}{}^{ab}R_{\rho \sigma}{}^{cd}M_{ab}M_{cd}D+\
+i\epsilon_{abcd}(R_{\mu \nu}{}^{ab}R_{\rho \sigma}{}^{c} K^d D - \\-R_{\mu \nu}{}^{ab}\tilde{R}_{\rho \sigma}{}^{c}P^{d}D)+(\frac{1}{2}\tilde{R}_{\mu \nu}{}^{a}R_{\rho \sigma} + 2\tilde{R}_{\mu \nu}{}^{a}R_{\rho \sigma}{}^{b})M_{ab}
+(\frac{1}{4}R_{\mu \nu}R_{\rho \sigma}- 2\tilde{R}_{\mu \nu}{}^{a}R_{\rho \sigma a})D
\Big].
\end{gathered}
\end{equation}
When evaluating the trace, we use:
\begin{equation}
\begin{gathered}
\operatorname{tr}[K^{d}D]=\operatorname{tr}[P^{d}D]=\operatorname{tr}[M_{ab}]= \operatorname{tr}[D]=0, \
\text{and}\quad \operatorname{tr}[M_{ab}M_{cd}D]=-\frac{1}{2}\epsilon_{abcd},
\end{gathered}
\end{equation}
This gives rise to the broken-symmetry action with only Lorentz invariance remaining:
\begin{equation}
\label{BrokenActionConformal}
S_{\mathrm{SO}(1,3)}=\frac{a_{CG}}{4}\int d^4x \epsilon^{\mu \nu \rho \sigma}\epsilon_{abcd}R_{\mu \nu}{}^{ab}R_{\rho \sigma}{}^{cd},
\end{equation}
Notably, $\tilde{a}_\mu$ no longer appears in the resulting expression. Consequently, we may set $\tilde{a}_\mu = 0$, simplifying the expressions for $\tilde{R}_{\mu \nu}{}^a$ and $R_{\mu \nu}{}^a$:
\begin{equation}
\begin{aligned}
\tilde{R}_{\mu \nu}{}^a & =mT_{\mu \nu}^{(0) a}(e)-2 m^2\tilde{a}_{[\mu} e_{\nu]}{}^a \longrightarrow mT_{\mu \nu}^{(0) a}(e), \\
R_{\mu \nu}{}^a &=mT_{\mu \nu}^{(0) a}(b)+2m^2 \tilde{a}_{[\mu} b_{\nu]}{}^a \longrightarrow mT_{\mu \nu}^{(0) a}(b).
\end{aligned}
\end{equation}
Since these components of the field strength tensor are not present in the action, we may set $\tilde{R}_{\mu \nu}{}^a = R_{\mu \nu}{}^a = 0$, choosing a torsion-free theory. Similarly, the absence of $R_{\mu \nu}$ in the action allows us to impose $R_{\mu \nu}=0$. Using its definition from \eqref{curves}, we then deduce a constraint linking the gauge fields $e$ and $b$:
\begin{equation}\label{ef}
e_\mu{}^a b_{\nu a}-e_{\nu}{}^{a}b_{\mu a}=0.
\end{equation}
This constraint admits multiple solutions, two of which are examined below.

\subsubsection*{Case $b_\mu{}^a = ae_\mu{}^a$}

This particular identification between $b$ and $e$, first suggested in \cite{Chamseddine:2002fd}, when substituted into the equation \eqref{BrokenActionConformal}, the following action is being retrieved:
\begin{equation}
\begin{aligned}
S_{\mathrm{SO}(1,3)} =\frac{a_{CG}}{4}\int d^4 x \epsilon^{\mu \nu \rho \sigma} \epsilon_{a b c d}&\left[R_{\mu \nu}^{(0) a b}-4m^2a\left(e_\mu{}^a e_\nu{}^b-e_\mu{}^b e_\nu{}^a\right)\right]\\
&\left[R_{\rho \sigma}^{(0) c d}-4m^2a\left(e_\rho{}^c e_\sigma{}^d-e_\rho{}^d e_\sigma{}^c\right)\right]
\end{aligned}
\end{equation}
Through straightforward algebraic manipulation, this action becomes:
\begin{equation}\label{so24finalaction}
\begin{aligned}
S_{\mathrm{SO}(1,3)}=\frac{a_{CG}}{4}\int d^4 x \epsilon^{\mu \nu \rho \sigma} \epsilon_{a b c d}[R_{\mu \nu}^{(0) a b}R_{\rho \sigma}^{(0) c d}-16m^2aR_{\mu \nu}^{(0) a b}e_\rho{}^c e_\sigma{}^d+\\
+64m^4a^2 e_\mu{}^a e_\nu{}^b e_\rho{}^c e_\sigma{}^d].
\end{aligned}
\end{equation}
This resulting action consists of three terms: one topological Gauss-Bonnet invariant (which does not affect field equations), the Einstein-Hilbert (Palatini) action equivalent in the vierbein formalism, and a term that behaves as a cosmological constant, respectively. For $a < 0$, the theory describes General Relativity in an Anti-de Sitter spacetime background.

\subsubsection*{Case $b_\mu{}^a = -\frac{1}{4} (R_\mu{}^a - \frac{1}{6} R e_\mu{}^a)$}

This relation between $b$ and $e$ is used in \cite{Kaku:1978nz} and \cite{freedman_vanproeyen_2012}, and when employed leads to the following action:
\begin{equation}
    \begin{aligned}
             S_W=\frac{a_{CG}}{4}&\int d^4 x \epsilon^{\mu \nu \rho \sigma} \epsilon_{a b c d}
             \\&\left[R_{\mu \nu}^{(0) a b}+\frac{1}{2}\left(m e_\mu{}^{[a} R_\nu{}^{b]}-me_\nu{}^{[a} R_\mu{}^{b]}\right)-\frac{1}{3} m^2 R e_\mu{}^{[a} e_\nu{}^{b]}\right]\\
             &\left[R_{\rho \sigma}^{(0) c d}+\frac{1}{2}\left(m e_\rho{}^{[c} R_\sigma{}^{d]}-m e_\sigma{}^{[c} R_\rho{}^{d]}\right)-\frac{1}{3} m^2 R e_\rho{}^{[c} e_\sigma{}^{d]}\right].
    \end{aligned}
\end{equation}

By introducing the rescaled vierbein $\tilde{e}_\mu{}^a = m e_\mu{}^a$ and using the antisymmetric property of the curvature 2-form, $R_{\mu \nu}^{(0)ab} = -R_{\nu \mu}^{(0)ab}$, we obtain:
\begin{equation}
    \begin{aligned}
             S_W=\frac{a_{CG}}{4}\int d^4 x&\epsilon^{\mu \nu \rho \sigma} \epsilon_{a b c d}\\&\left[R_{\mu \nu}^{(0) a b}-\frac{1}{2}\left(\tilde{e}_\mu{}^{[a} R_\nu{}^{b]}-\tilde{e}_\nu{}^{[a} R_\mu{}^{b]}\right)+\frac{1}{3} R \tilde{e}_\mu{}^{[a} \tilde{e}_\nu{}^{b]}\right]\\
             &\left[R_{\rho \sigma}^{(0) c d}-\frac{1}{2}\left( \tilde{e}_\rho{}^{[c} R_\sigma{}^{d]}- \tilde{e}_\sigma{}^{[c} R_\rho{}^{d]}\right)+\frac{1}{3} R \tilde{e}_\rho{}^{[c} \tilde{e}_\sigma{}^{d]}\right].
    \end{aligned}
\end{equation}

Each bracket matches the form of the Weyl conformal tensor $C_{\mu \nu}{}^{a b}$, which means that the above action is actually the following:
\begin{align}
  S_W=\frac{a_{CG}}{4}\int d^4 x \epsilon^{\mu \nu \rho \sigma} \epsilon_{a b c d}C_{\mu \nu}{}^{a b}C_{\rho \sigma}{}^{c d},
\end{align}

This corresponds to the well-known scale-invariant Weyl action in four dimensions:
\begin{equation}
S_W =2a_{CG}\int \mathrm{d}^4 x\left(R_{\mu \nu} R^{\nu \mu}-\frac{1}{3} R^2\right).
\end{equation}

\subsection{Spontaneous symmetry breaking by introducing two scalars in vector representations}

The spontaneous symmetry breaking (SSB) of the $SO(2,4)$ gauge symmetry can be realized by employing two scalar fields in the vector representation of $SO(6)$, noting that the Lie algebra of $SO(2,4)$ is isomorphic to those of $SU(4)$ and $SO(6)$. Consequently, the $SO(2,4)$ gauge group may be spontaneously broken to $Sp_4 = SO(5)$ by introducing a scalar field in the $6$-dimensional vector representation of $SO(6)$, which acquires a vacuum expectation value (vev) in the $\langle 1 \rangle$ component, in accordance with the branching rules of the $SO(6)$ representation under its maximal subgroup, $SO(5)$:
\begin{equation}
\begin{aligned}
SO(6) &\supset SO(5) \\
6 &= 1 + 5.
\end{aligned}
\end{equation}
Next, the $SO(5)$ symmetry is spontaneously broken by another scalar field in the vector representation $5$, with the branching rule:
\begin{equation}
\begin{aligned}
SO(5) &\supset SU(2) \times SU(2) \\
5 &= (1,1) + (2,2).
\end{aligned}
\end{equation}
Since the algebra of $SU(2) \times SU(2)$ is isomorphic to both $SO(4)$ and $SO(1,3)$, breaking the symmetry with a scalar field in the $5$ of $SO(5)$ that acquires a vev in the $\langle 1,1 \rangle$ component yields the desired $SO(1,3)$ gauge group. Let us now turn to the generators of the $SO(2,4)$ algebra:
\begin{equation}
\left[ J_{AB}, J_{CD} \right] = \eta_{BC} J_{AD} + \eta_{AD} J_{BC} - \eta_{AC} J_{BD} - \eta_{BD} J_{AC}.
\end{equation}

Here, $A, B = 1, \dots, 6$, and the signature of the metric is $\eta_{AB} = (-1, 1, 1, 1, -1, 1)$. This choice aligns with the two-step symmetry breaking procedure: first breaking a spacelike direction (the 6th component), then a timelike one (the 5th component). A single gauge field $A_\mu{}^{AB}$ represents all generators and is written as $A_\mu = \frac{1}{2} A_\mu{}^{AB} J_{AB}$. Thus, the field strength tensor $F_{\mu \nu} = \frac{1}{2} F_{\mu \nu}{}^{AB} J_{AB}$ is given by:
\begin{equation}
F_{\mu \nu} = [D_\mu, D_\nu] = \partial_\mu A_\nu - \partial_\nu A_\mu + [A_\mu, A_\nu] \rightarrow
\end{equation}
\begin{equation}
F_{\mu \nu}{}^{AB} = \partial_\mu A_\nu{}^{AB} - \partial_\nu A_\mu{}^{AB} + A_\mu{}^A{}_C A_\nu{}^{CB} - A_\nu{}^A{}_C A_\mu{}^{CB}.
\end{equation}

To build an $SO(2,4)$-invariant quadratic action, we introduce two scalar fields $\phi^E$ and $\chi^F$, each in the vector representation $6$ of $SO(2,4)$, and two mass parameters $m_\phi$ and $m_\chi$:
\begin{equation}
\begin{aligned}
S_{SO(2,4)} = a_{CG} \int d^4x [&\epsilon^{\mu \nu \rho \sigma} \epsilon_{ABCDEF} \phi^E \chi^F m_\phi m_\chi \frac{1}{4} F_{\mu \nu}{}^{AB} F_{\rho \sigma}{}^{CD} + \\
&+ \lambda_\phi (\phi^E \phi_E - m_\phi^{-2}) + \lambda_\chi (\chi^F \chi_F + m_\chi^{-2})],
\end{aligned}
\end{equation}
where $a_{CG}$ is dimensionless, and $m_\phi \geq m_\chi$. In our chosen gauge, the auxiliary scalar field $\phi^E$ is
\begin{equation}
\phi^E = \phi^0 = (0,0,0,0,0, m_\phi^{-1}),
\end{equation}
with $m_\phi^2 > 0$ since $\phi^0$ is spacelike. Then, the component gauge field for the five broken generators becomes:
\begin{equation}
A_\mu{}^{j6} = m_\phi f_\mu{}^j,
\end{equation}
where $f$ denotes the rescaled gauge field.

The field strength tensor now reads:
\begin{equation}
F_{\mu \nu}{}^{jk} = \partial_\mu A_\nu{}^{jk} - \partial_\nu A_\mu{}^{jk} + A_\mu{}^j{}_l A_\nu{}^{lk} - A_\nu{}^j{}_l A_\mu{}^{lk} - m_\phi^2 (f_\mu{}^j f_\nu{}^k - f_\mu{}^k f_\nu{}^j),
\end{equation}
for $j,k = 1, \dots, 5$, and the resulting $SO(2,3)$-symmetric action becomes:
\begin{equation}
S_{SO(2,3)} = a_{CG} \int d^4x \left[ \epsilon^{\mu \nu \rho \sigma} \epsilon_{ijklm} m_\chi \chi^m \frac{1}{4} F_{\mu \nu}{}^{ij} F_{\rho \sigma}{}^{kl} + \lambda_\chi (\chi^m \chi_m + m_\chi^{-2}) \right].
\end{equation}
Next, we perform the second SSB by fixing the second auxiliary scalar field in the gauge:
\begin{equation}
\chi^m = \chi^0 = (0,0,0,0, m_\chi^{-1}),
\end{equation}
with $m_\chi^2 > 0$, as $\chi^0$ is timelike.

Recall that the conformal group $SO(2,4)$ has 15 generators, interpreted in four dimensions as six Lorentz generators $M_{ab}$, four translations $P_a$, four conformal boosts $K_a$, and one dilatation $D$. The respective gauge fields are $\omega_\mu{}^{ab}$, $e_\mu{}^a$, $b_\mu{}^a$, and $\tilde{a}_\mu$. In contrast, $SO(2,3)$ includes only the six Lorentz and four translation generators.

Thus, the five broken generators in $SO(2,4) \rightarrow SO(2,3)$ are linear combinations of $P_a$, $K_a$, and $D$.

The field $f_\mu{}^j$ in 4D notation is:
\begin{equation}
f_\mu{}^j = \big(b_\mu{}^a - e_\mu{}^a, -\tilde{a}_\mu \big),
\end{equation}
for $j = 1, \dots, 5$ and $a,b = 1, \dots, 4$. The second symmetry breaking yields:
\begin{equation}
A_\mu{}^{a5} = - m_\chi (b_\mu{}^a + e_\mu{}^a).
\end{equation}

Hence, four more generators are broken, leaving only the six Lorentz generators $M_{ab}$. The full set of gauge fields becomes:
\begin{align}
A_\mu{}^{j6} = m_\phi f_\mu{}^j \Rightarrow \begin{cases}
A_\mu{}^{a6} = m_\phi (b_\mu{}^a - e_\mu{}^a) \\
A_\mu{}^{56} = - m_\phi \tilde{a}_\mu
\end{cases},\quad
A_\mu{}^{a5} = - m_\chi (b_\mu{}^a + e_\mu{}^a),
\end{align}
and the remaining unbroken gauge fields correspond to the spin connection:
\begin{equation}
A_\mu{}^{ab} = \omega_\mu{}^{ab}.
\end{equation}

This leads us to:
\begin{align}\label{curv2}
F_{\mu \nu}{}^{a b}= &\partial_\mu \omega_\nu{}^{a b}-\partial_\nu \omega_\mu{}^{a b}-\omega_\mu{}^{a c} \omega_{\nu c}{}^b+\omega_\nu{}^{a c} \omega_{\mu c}{}^b \nonumber\\
&+( {m_\chi}^2-{m_\phi}^2)\left(e_\mu{}^a e_\nu{}^b-e_\nu{}^a e_\mu{}^b+b_\mu{}^a b_\nu{}^b-b_\nu{}^a b_\mu{}^b\right)\nonumber\\
&-( {m_\chi}^2+{m_\phi}^2)(b_\mu{}^a e_\nu{}^b-b_\nu{}^a e_\mu{}^b+e_\mu{}^a b_\nu{}^b-e_\nu{}^a b_\mu{}^b)\longrightarrow\nonumber\\
F_{\mu \nu}{}^{a b}=&R_{\mu \nu}{}^{a b}+( {m_\chi}^2-{m_\phi}^2)\left(e_\mu{}^a e_\nu{}^b-e_\nu{}^a e_\mu{}^b+b_\mu{}^a b_\nu{}^b-b_\nu{}^a b_\mu{}^b\right)\\
&-( {m_\chi}^2+{m_\phi}^2)(b_\mu{}^a e_\nu{}^b-b_\nu{}^a e_\mu{}^b+e_\mu{}^a b_\nu{}^b-e_\nu{}^a b_\mu{}^b),\nonumber
\end{align}
and
\begin{align}
\label{tors2}
F_{\mu \nu}{}^{a 5}&=-{m_\chi} (\partial_\mu e_\nu{}^a-\partial_\nu e_\mu{}^a-\omega_\mu{}^{a b} e_{\nu b}+\omega_\nu{}^{a b} e_{\mu b}+\partial_\mu b_\nu{}^a-\partial_\nu b_\mu{}^a-\omega_\mu{}^{a b} b_{\nu b}+\omega_\nu{}^{a b} b_{\mu b}) + \nonumber \\
&+{m_\phi}^2[\tilde{a}_\mu( e_\nu{}^a -b_\nu{}^a)-\tilde{a}_\nu( e_\mu{}^a-b_\mu{}^a)]\longrightarrow \nonumber\\
F_{\mu \nu}{}^{a 5}&=-{m_\chi}[ T_{\mu \nu}{}^a(e)+T_{\mu \nu}{}^a(b)] +{m_\phi}^2[\tilde{a}_\mu( e_\nu{}^a -b_\nu{}^a)-\tilde{a}_\nu( e_\mu{}^a-b_\mu{}^a)].
\end{align}

Since $F_{\mu \nu}{}^{a5}$ does not appear in the final action, we can consistently set $F_{\mu \nu}{}^{a5} = 0$ and hence $\tilde{a}_\mu = 0$, which may imply a torsion-free theory.

The final broken symmetry-bearing action is:
\begin{equation}
\begin{aligned}\label{act2}
    S_{SO(1,3)} =\frac{a_{CG}}{4}\int &d^4x \epsilon^{\mu \nu \rho \sigma}\epsilon_{abcd}F_{\mu \nu}{ }^{a b} F_{\rho \sigma}{ }^{c d}= \\
    =\frac{a_{CG}}{4}\int & d^4x \epsilon^{\mu \nu \rho \sigma}\epsilon_{abcd}[R_{\mu \nu}{ }^{a b} R_{\rho \sigma}{ }^{c d} +\\
    &+2( {m_\chi}^2-{m_\phi}^2)R_{\mu \nu}{ }^{a b}(e_\rho{}^c e_\sigma{}^d-e_\sigma{}^c e_\rho{}^d+b_\rho{}^c b_\sigma{}^d-b_\sigma{}^c b_\rho{}^d)\\
    &-2( {m_\chi}^2+{m_\phi}^2)R_{\mu \nu}{ }^{a b}(b_\rho{}^c e_\sigma{}^d-b_\sigma{}^c e_\rho{}^d+e_\rho{}^c b_\sigma{}^d-e_\sigma{}^c b_\rho{}^d)\\
 &-2( {m_\chi}^4-{m_\phi}^4)\left(e_\mu{}^a e_\nu{}^b-e_\nu{}^a e_\mu{}^b+b_\mu{}^a b_\nu{}^b-b_\nu{}^a b_\mu{}^b\right) \\
&\qquad\qquad\qquad\qquad\times(b_\rho{}^c e_\sigma{}^d-b_\sigma{}^c e_\rho{}^d+e_\rho{}^c b_\sigma{}^d-e_\sigma{}^c b_\rho{}^d)\\
    &+( {m_\chi}^2-{m_\phi}^2)^2\left(e_\mu{}^a e_\nu{}^b-e_\nu{}^a e_\mu{}^b+b_\mu{}^a b_\nu{}^b-b_\nu{}^a b_\mu{}^b\right)\\
    &\qquad\qquad\qquad\qquad\times(e_\rho{}^c e_\sigma{}^d-e_\sigma{}^c e_\rho{}^d+b_\rho{}^c b_\sigma{}^d-b_\sigma{}^c b_\rho{}^d)\\
    & +( {m_\chi}^2+{m_\phi}^2)^2\left(b_\mu{}^a e_\nu{}^b-b_\nu{}^a e_\mu{}^b+e_\mu{}^a b_\nu{}^b-e_\nu{}^a b_\mu{}^b\right)\\
    &\qquad\qquad\qquad\qquad\times(b_\rho{}^c e_\sigma{}^d-b_\sigma{}^c e_\rho{}^d+e_\rho{}^c b_\sigma{}^d-e_\sigma{}^c b_\rho{}^d)].
\end{aligned}
\end{equation}

Furthermore, the component $F_{\mu \nu}{}^{56}$ is also absent, so we set:
\begin{equation}
F_{\mu \nu}{}^{56} = m_\phi [\partial_\mu \tilde{a}_\nu - \partial_\nu \tilde{a}_\mu - m_\chi (e_{\mu a} b_\nu{}^a - e_{\nu a} b_\mu{}^a)] = 0.
\end{equation}
With $\tilde{a}_\mu = 0$, this implies:
\begin{equation}\label{efa}
e_{\mu a} b_\nu{}^a - e_{\nu a} b_\mu{}^a = 0.
\end{equation}

\subsubsection*{Case $b_\mu{}^a=ae_\mu{}^a$}\label{casea}
As shown previously, in this case the final action becomes
\begin{equation}
\begin{aligned}
\label{34}
 S_{SO(1,3)} =\frac{a_{CG}}{4}\int d^4x &\epsilon^{\mu \nu \rho \sigma}\epsilon_{abcd}F_{\mu \nu}{ }^{a b} F_{\rho \sigma}{ }^{c d}= \\
    =\frac{a_{CG}}{4}\int d^4x \epsilon^{\mu \nu \rho \sigma}\epsilon_{abcd} & \Big[R_{\mu \nu}{ }^{a b} R_{\rho \sigma}{ }^{c d} +4\Big({m_\chi}^2(1-a)^2-{m_\phi}^2(1+a)^2\Big)R_{\mu \nu}{ }^{a b}e_\rho{}^c e_\sigma{}^d\\
    & +4\Big({m_\chi}^2(1-a)^2-{m_\phi}^2(1+a)^2\Big)^2 e_\mu{}^a e_\nu{}^b e_\rho{}^c e_\sigma{}^d\Big]
\end{aligned}
\end{equation}

As before, the first term of the action \eqref{34} is a G-B topological term, the second term is the E-H (Palatini) action, while the last term is the cosmological constant. When ${m_\chi}^2/{m_\phi}^2>(1+a)^2/(1-a)^2$, the above action describes GR in AdS space.

When $m_\phi=m_\chi\equiv m$, we obtain:
\begin{equation}
\begin{aligned}
\label{35}
  S_{SO(1,3)} =\frac{a_{CG}}{4}\int &d^4x \epsilon^{\mu \nu \rho \sigma}\epsilon_{abcd}F_{\mu \nu}{ }^{a b} F_{\rho \sigma}{ }^{c d} =\\
    =\frac{a_{CG}}{4}\int & d^4x \epsilon^{\mu \nu \rho \sigma}\epsilon_{abcd}\Big[R_{\mu \nu}{ }^{a b} R_{\rho \sigma}{ }^{c d} -16{m}^2aR_{\mu \nu}{ }^{a b} e_\rho{}^c e_\sigma{}^d
    +64{m}^4 a^2e_\mu{}^a e_\nu{}^b e_\rho{}^c e_\sigma{}^d\Big],
\end{aligned}
\end{equation}
which, when $a<0$ describes GR in AdS space, and is completely equivalent to the action resulting in the case of the SSB with a scalar in the adjoint rep, analysed above.

\subsubsection*{Case $b_\mu{}^a=-\frac{1}{4}(R_\mu{}^a+\frac{1}{6}R e_\mu{}^a)$ and $m_\phi=m_\chi$}
\label{caseb}

Choosing this relation among the gauge fields $e$ and $b$, we obtain again the following action:
\begin{equation}
    \begin{aligned}
             S=\frac{a_{CG}}{4}\int d^4 x \epsilon^{\mu \nu \rho \sigma} \epsilon_{a b c d}&\left[R_{\mu \nu}{}^{a b}+\frac{1}{2}\left(m e_\mu{}^{[a} R_\nu{}^{b]}-me_\nu{}^{[a} R_\mu{}^{b]}\right)-\frac{1}{3} m^2 R e_\mu{}^{[a} e_\nu{}^{b]}\right]\\
             &\left[R_{\rho \sigma}{}^{c d}+\frac{1}{2}\left(m e_\rho{}^{[c} R_\sigma{}^{d]}-m e_\sigma{}^{[c} R_\rho{}^{d]}\right)-\frac{1}{3} m^2 R e_\rho{}^{[c} e_\sigma{}^{d]}\right],
    \end{aligned}
\end{equation}
where $m \equiv m_\phi=m_\chi$. Applying the rescaled vierbein, $\tilde{e}_\mu{}^{a}=m e_\mu{}^{a}$, and the antisymmetric property of the curvature 2-form, $R_{\mu \nu}{}^{a b}=-R_{\nu \mu}{}^{a b}$, we again obtain:
\begin{equation}
    \begin{aligned}
             S=\frac{a_{CG}}{4}\int d^4 x \epsilon^{\mu \nu \rho \sigma} \epsilon_{a b c d}&\left[R_{\mu \nu}{}^{a b}-\frac{1}{2}\left(\tilde{e}_\mu{}^{[a} R_\nu{}^{b]}-\tilde{e}_\nu{}^{[a} R_\mu{}^{b]}\right)+\frac{1}{3} R \tilde{e}_\mu{}^{[a} \tilde{e}_\nu{}^{b]}\right]\\
             &\left[R_{\rho \sigma}{}^{c d}-\frac{1}{2}\left( \tilde{e}_\rho{}^{[c} R_\sigma{}^{d]}- \tilde{e}_\sigma{}^{[c} R_\rho{}^{d]}\right)+\frac{1}{3} R \tilde{e}_\rho{}^{[c} \tilde{e}_\sigma{}^{d]}\right].
    \end{aligned}
\end{equation}
The above action is equivalent to:
\begin{align}
\label{weyl1}
             S=\frac{a_{CG}}{4}\int d^4 x \epsilon^{\mu \nu \rho \sigma} \epsilon_{a b c d}C_{\mu \nu}{}^{a b}C_{\rho \sigma}{}^{c d},
\end{align}
where $C_{\mu \nu}{}^{a b}$ is the Weyl conformal tensor. This action, as discussed before, actually leads to
\begin{equation}\label{weyl2}
S =2 a_{CG}\int \mathrm{d}^4 x\left(R_{\mu \nu} R^{\nu \mu}-\frac{1}{3} R^2\right),
\end{equation}
which describes the four-dimensional scale invariant Weyl theory of gravity.

The Weyl action of WG in the forms given in eqs. \eqref{weyl1} and \eqref{weyl2}, being scale invariant, naturally does not contain a cosmological constant. WG is an attractive possibility for describing gravity at high scales (for some recent developments see \cite{Maldacena:2011mk, mannheim, Anastasiou:2016jix, ghilencea2023, Hell:2023rbf, Condeescu:2023izl}, as CG does. However, in the case that WG is obtained after the SSB of the CG, as is described above, a question remains on how one can obtain Einstein gravity from the SSB of WG. Our suggestion is the following. We start again from CG and we introduce a scalar in the \textbf{15} rep of $SU(4) \sim SO(6) \sim SO(2,4)$ as described above and by choosing the relation
\begin{equation}
\label{b-e_relation}
b_\mu{}^a=-\frac{1}{4}\left(R_\mu{}^a-\frac{1}{6} R e_\mu{}^a\right),
\end{equation}
we are led after the SSB of the scalar $\textbf{15}$-plet to the Weyl action. In addition, we introduce a scalar in the 2nd rank anti-symmetric tensor of $SU(4)$, \textbf{6}, which after SSB leads to the E-H action. One can easily see the result of these breakings by considering the decompositions of the \textbf{15} generators of $SU(4)$ under $SU(2) \times SU(2) \times U(1)$, describing the gauge group of the Lorentz group and dilatations to which $SU(4)$ breaks after the SSB due to the scalar $\textbf{15}$-plet,
\begin{align}
\label{SU4ToSU2SU2U1}
    SU(4)&\xrightarrow{<\textbf{15}>}SU(2) \times SU(2) \times U(1) \nonumber\\
    \textbf{15}&=[(\textbf{3},\textbf{1})_0+(\textbf{1},\textbf{3})_0]+(\textbf{1},\textbf{1})_0+(\textbf{2},\textbf{2})_2+(\textbf{2},\textbf{2})_2\,,
\end{align}
where the $[(\textbf{3},\textbf{1})_0+(\textbf{1},\textbf{3})_0]$ describes the generators of the Lorentz gauge group, $M_{ab}$, the $(\textbf{1},\textbf{1})_0$ the generator of dilatations, $D$, the $(\textbf{2},\textbf{2})_2$ the generators of the translations, $P_a$ and the $(\textbf{2},\textbf{2})_2$ the generators of conformal transformations, $K_a$ (there is an arbitrariness in the choice of the last two sets of generators). The generators $P_a$ and $K_a$ are broken due to the SSB of the scalar $\textbf{15}$-plet. Similarly, the decomposition of the 15 generators of $SU(4)$ under the $SO(5)$ to which it breaks after the SSB of the scalar $\textbf{6}$-plet is,
\begin{align}
\label{SU4ToSO5}
    SU(4)&\xrightarrow{<\textbf{6}>}SO(5) \nonumber\\
    \textbf{15}&=\textbf{10}+\textbf{5}\,,
\end{align}
where the \textbf{10} describes the generators of the unbroken gauge group $SO(5)$ and \textbf{5} the broken generators. To identify the unbroken and the broken generators in \eqref{SU4ToSO5} it helps to consider the decomposition of reps \textbf{10} and \textbf{5} of $SO(5)$ under the $SU(2) \times SU(2)$ describing the Lorentz gauge group in \eqref{SU4ToSU2SU2U1},
\begin{align}
    SO(5)& \supset SU(2) \times SU(2) \nonumber\\
    \textbf{10}&=(\textbf{3},\textbf{1})+(\textbf{1},\textbf{3})+(\textbf{1},\textbf{1})+(\textbf{2},\textbf{2})\,,\\
    \textbf{5}&=(\textbf{1},\textbf{1})+(\textbf{2},\textbf{2})\,.
\end{align}
Now one can easily recognize the ten unbroken generators from the SSB of the scalar \textbf{6}-plet correspond to the Lorentz generators, $M_{ab}$ and the generators of the translations, $P_a$ (which, though they were broken by the $<$\textbf{15}$>$), while the five broken generators can be identified with the generators $(\textbf{1},\textbf{1})$ of dilatations and the $(\textbf{2},\textbf{2})$ of $K_a$.

In summary, $<$\textbf{15}$>$ breaks the generators of $P_a$ and $K_a$, leaving unbroken the Lorentz rotation generators, $M_{ab}$ and the dilaton generator, $D$, while $<$\textbf{6}$>$ breaks the dilaton generator, $D$ and gives an additional contribution to the breaking of the generators $K_a$ (and to the masses of the corresponding gauge bosons).

It should be noted, as it was discussed explicitly in \cite{Patellis:2025qbl} that the vanishing of the two torsions $\tilde{R}_{\mu\nu}{}^a$, and $R_{\mu\nu}{}^a$ as well as the curvature tensor, $F_{\mu\nu}$ which is satisfied on-shell guarantee the equivalence of the diffeomorphisms and gauge transformations. At this point, it should also be noted that the relation $b_\mu{}^a = -\frac{1}{4}(R_\mu{}^a - \frac{1}{6}R e_\mu{}^a)$, which eventually leads to WG, is an additional constraint and is not necessary in order to guarantee the equivalence of the diffeomorphisms and gauge transformations. Furthermore, it is maybe worth noting that the gauge fixing used for the spontaneous symmetry breaking is a general feature of such theories and does not restrict in any way the formulation of gravities as gauge theories.

\section{Noncommutative (Fuzzy) Gravity}

\subsection{Gauge Theories on Noncommutative Spaces}

We begin by reviewing the essential ingredients for constructing gauge theories on noncommutative (NC) spaces, which form the foundation of our approach. In the framework of NC geometry, gauge fields arise naturally and are closely associated with the notion of the covariant coordinate~\cite{Madore:2000en}. This concept plays a role analogous to that of the covariant derivative in conventional gauge theory, as we shall see.

Consider a scalar field $\phi(X_a)$ defined on a fuzzy space, where the coordinates $X_a$ obey non-trivial commutation relations. The field $\phi$ transforms under a representation of a gauge group $G$, and an infinitesimal gauge transformation with parameter $\lambda(X_a)$ acts as:
\begin{equation}
\delta \phi(X) = \lambda(X) \phi(X)\,.
\end{equation}
When the gauge parameter $\lambda(X)$ is a scalar function of the coordinates, the transformation is Abelian, and $G = U(1)$. If instead $\lambda(X)$ is a $P \times P$ Hermitian matrix function, then the gauge transformation is non-Abelian, corresponding to $G = U(P)$.

Importantly, the NC coordinates $X_\alpha$ are themselves taken to be invariant under gauge transformations:
\begin{equation}
\delta X_\alpha = 0\,.
\end{equation}
However, applying a gauge transformation to the product of a coordinate and a field yields:
\begin{equation}
\delta(X_a \phi) = X_a \lambda(X) \phi\,,
\end{equation}
which is not covariant in general, since:
\begin{equation}
X_a \lambda(X) \phi \neq \lambda(X) X_a \phi\,.
\end{equation}

To restore covariance, we adopt an approach analogous to that of conventional gauge theories, where covariant derivatives ensure the appropriate transformation behavior. This motivates the introduction of the \emph{covariant coordinate} $\mathcal{X}_a$, which transforms as:
\begin{equation}
\label{3.5}
\delta(\mathcal{X}_a \phi) = \lambda\, \mathcal{X}_a \phi\,,
\end{equation}
provided that
\begin{equation}
\label{3.6}
\delta \mathcal{X}_a = [\lambda, \mathcal{X}_a]\,.
\end{equation}
This condition is satisfied by defining the covariant coordinate as:
\begin{equation}
\mathcal{X}_a \equiv X_a + A_a\,,
\end{equation}
where $A_a$ is interpreted as the gauge connection of the theory. From equations \eqref{3.5} and \eqref{3.6}, the gauge transformation of $A_a$ follows as:
\begin{equation}
\delta A_a = -[X_a, \lambda] + [\lambda, A_a]\,,
\end{equation}
confirming its role as the gauge connection\footnote{For further discussion, see \cite{Aschieri:2005wm, Aschieri:2004vh}.}.

The field strength associated with $A_a$ is accordingly defined as:
\begin{equation}
F_{ab} \equiv [X_a, A_b] - [X_b, A_a] + [A_a, A_b] - C_{ab}{}^c A_c = [\mathcal{X}_a, \mathcal{X}_b] - C_{ab}{}^c \mathcal{X}_c\,,
\end{equation}
and transforms covariantly under gauge transformations:
\begin{equation}
\delta F_{ab} = [\lambda, F_{ab}]\,.
\end{equation}

\subsection{The Background Space}
\label{sec3.2}

Before developing the full gauge theory of fuzzy gravity (FG), we first have to define the background spacetime. In this work, we consider both four-dimensional de Sitter ($dS_4$) and anti-de Sitter ($AdS_4$) spaces.

Both geometries can be realized as embedded in a five-dimensional flat space, subject to the constraint:
\begin{equation}
\label{eq:constraint}
\eta_{\mu \nu} x^\mu x^\nu = s R^2\,,
\end{equation}
where $R$ denotes the curvature radius, $\eta_{\mu\nu} = \mathrm{diag}(-1, 1, 1, 1, s)$ is the ambient-space metric, and $s = \pm 1$ determines the signature: $s = +1$ corresponds to $dS_4$ with isometry group $SO(1,4)$, while $s = -1$ corresponds to $AdS_4$ with isometry group $SO(2,3)$. Throughout this section, we treat both cases simultaneously.

These isometry groups contain 10 generators, which is insufficient to fully covariantize both the 5 embedding coordinates and the local Lorentz symmetry. We thus consider the minimal covariant extensions of these groups: $SO(1,5)$ in the de Sitter case and $SO(2,4)$ for anti-de Sitter. Each of these has 15 generators, capable of accommodating the 5 coordinates as well as the 10 generators needed to form a Lorentz subgroup.

To further explore the underlying structure, and further extend the algebra of observables, we consider an even larger group, $SO(1,6)$ or $SO(2,5)$,\footnote{This extension is inspired by the logic of Snyder and Yang, as discussed in Appendix~\ref{appA}.} containing 21 generators. This leads to the following symmetry breaking chain:
\begin{align}
SO(1,6) &\supset SO(1,5) \supset SO(1,4) \supset SO(1,3)\,, \\
\text{or } SO(2,5) &\supset SO(2,4) \supset SO(2,3) \supset SO(1,3)\,.
\end{align}

Let us denote the generators of $SO(1,6)$ by $J_{MN}$, with $M, N = 0, \dots, 6$. These obey the standard Lorentz-like algebra:
\begin{equation}
[J_{MN}, J_{RS}] = i \left( \eta_{MR} J_{NS} + \eta_{NS} J_{MR} - \eta_{NR} J_{MS} - \eta_{MS} J_{NR} \right)\,,
\end{equation}
with $\eta_{MN} = \mathrm{diag}(-1, 1, 1, 1, s, 1, 1)$.

Following the above decompositions form the initial groups down to $SO(1,3)$, we obtain the following nontrivial commutation relations:
\begin{equation}
\begin{aligned}
&[J_{ij}, J_{kl}] = i(\eta_{ik}J_{jl} + \eta_{jl}J_{ik} - \eta_{jk}J_{il} - \eta_{il}J_{jk})\,, \\
&[J_{ij}, J_{k6}] = i(\eta_{ik}J_{j6} - \eta_{jk}J_{i6})\,, \qquad [J_{ij}, J_{k5}] = i(\eta_{ik}J_{j5} - \eta_{jk}J_{i5})\,, \\
&[J_{ij}, J_{k4}] = i(\eta_{ik}J_{j4} - \eta_{jk}J_{i4})\,, \qquad [J_{i6}, J_{j6}] = i J_{ij}\,, \\
&[J_{i6}, J_{j5}] = -i \eta_{ij} J_{56}\,, \qquad [J_{i6}, J_{j4}] = -i \eta_{ij} J_{46}\,, \\
&[J_{i6}, J_{56}] = i J_{i5}\,, \qquad [J_{i6}, J_{46}] = i J_{i4}\,, \\
&[J_{i5}, J_{j5}] = i J_{ij}\,, \qquad [J_{i5}, J_{j4}] = -i \eta_{ij} J_{45}\,, \\
&[J_{i5}, J_{56}] = -i J_{i6}\,, \qquad [J_{i5}, J_{45}] = i J_{i4}\,, \\
&[J_{i4}, J_{j4}] = i s J_{ij}\,, \qquad [J_{i4}, J_{46}] = -i s J_{i6}\,, \qquad [J_{i4}, J_{45}] = -i s J_{i5}\,, \\
&[J_{56}, J_{46}] = -i J_{45}\,, \qquad [J_{56}, J_{45}] = i J_{46}\,, \qquad [J_{46}, J_{45}] = -i s J_{56}\,.
\end{aligned}
\end{equation}
Here, the indices $i,j,k,l = 0,\dots,3$ denote four-dimensional spacetime directions, with $\eta_{ij} = \mathrm{diag}(-1,1,1,1)$.

We now define the relevant observables , i.e. the NC tensor, the coordinates and the momenta, in terms of these generators:
\begin{equation}
\Theta_{ij} = \hbar J_{ij}\,, \quad X_i = \lambda J_{i5}\,, \quad P_i = \frac{\hbar}{\lambda} J_{i4}\,,
\end{equation}
and the remaining operators as:
\begin{equation}
Q_i = \frac{\hbar}{\lambda} J_{i6}\,, \quad q = J_{56}\,, \quad p = J_{46}\,, \quad h = J_{45}\,.
\end{equation}
Given the identifications above, we are led to the following algebra:
\begin{equation}    
\begin{gathered}
\label{backgroundAlgebra}
    [\Theta_{ij}, \Theta_{kl}] = i\hbar(\eta_{ik}\Theta_{jl} + \eta_{jl}\Theta_{ik} - \eta_{jk}\Theta_{il} - \eta_{il}\Theta_{jk}), \\
    [\Theta_{ij}, X_k] = \frac{i}{\hbar} (\eta_{ik}X_j - \eta_{jk}X_i), \ 
    [\Theta_{ij}, P_k] = \frac{i}{\hbar} (\eta_{ik}P_j - \eta_{jk}P_i), \\
    [\Theta_{ij}, Q_k] = \frac{i}{\hbar} (\eta_{ik}Q_j - \eta_{jk}Q_i), \  [Q_i, Q_j] = i\frac{\hbar}{\lambda^2} \Theta_{ij}, \ 
    [X_i, X_j] = i\frac{\lambda^2}{\hbar} \Theta_{ij}, \\ 
    [P_i, P_j] = i s \frac{\hbar}{\lambda^2} \Theta_{ij}, \
    [Q_i, X_j] = -i \frac{\hbar}{\lambda^2} \eta_{ij} q, \ 
    [Q_i, P_j] = -i \frac{\hbar^2}{\lambda^2} \eta_{ij} p, \\
    [X_i, P_j] = -i\hbar \eta_{ij} h, \ 
    [Q_i, q] = i \frac{\hbar}{\lambda^2} X_i, \ 
    [Q_i, p] = i P_i, \\
    [X_i, q] = -i \frac{\lambda^2}{\hbar} Q_i, \ 
    [X_i, h] = i \frac{\lambda^2}{\hbar} P_i, \\
    [P_i, p] = -i s Q_i, \ 
    [P_i, h] = -i s \frac{\hbar}{\lambda^2} X_i, \\
    [q, p] = -i h, \ 
    [q, h] = i p, \ 
    [p, h] = -i s q.
\end{gathered}
\end{equation}

This algebra encodes more than just the  noncommutativity of the coordinates (as in Snyder's approach) and the momenta as well as Heisenberg-type relations among them (as in Yang’s framework). Moreover, it provides additional structure that will later be reflected in the choice of gauge group.

\subsection{Gauge group and representation}
To formulate a gauge theory of gravity on the spacetime we're considering, our first step is to pinpoint the appropriate symmetry group to be gauged. This will naturally be the isometry group of the background space, which is $SO(1,4)$ for de Sitter ($dS_4$) space, or $SO(2,3)$ for Anti-de Sitter ($AdS_4$) space.

However, when constructing gauge theories on noncommutative spaces, we must account for both the commutators and the anticommutators between various fields. To clarify this, let's take two arbitrary elements from a Lie algebra, $\varepsilon(X)=\varepsilon^{a}(X) T_{a}$ and $\phi(X)=\phi^{a}(X) T_{a}$, where $T_a$ are the algebra generators. Their commutator is expressed as:
\begin{equation}
\label{anticom}
[\varepsilon, \phi]=\frac{1}{2}\{\varepsilon^{a}, \phi^{b}\}\left[T_{a}, T_{b}\right]+\frac{1}{2}\left[\varepsilon^{a}, \phi^{b}\right]\{T_{a}, T_{b}\}.
\end{equation}
In standard commutative cases, the second term vanishes because the component functions $\varepsilon^a$ and $\phi^b$ commute. But on a noncommutative space, this term persists due to the non-trivial commutation relations of these components, meaning the anticommutator $\{T_a, T_b\}$ remains in the expression.

Crucially, the anticommutator of two generators doesn't generally belong to the original algebra, which means the algebra fails to close. To address this, we have two main options: either explicitly include every anticommutator that falls outside the initial gauge group as an element of the algebra (which would lead to an infinite-dimensional, universal enveloping algebra), or, as we choose to do, select a specific representation for the generators and expand the original gauge group to a larger symmetry group where the algebra is closed under both commutators and anticommutators.

Consequently, the gauge group is expanded from $SO(1,4)$ to $SO(1,5) \times U(1)$ for $dS_4$, and from $SO(2,3)$ to $SO(2,4) \times U(1)$ for $AdS_4$. It's worth noting that the algebra presented in equation \eqref{backgroundAlgebra} already captures the structure of $SO(1,5)$ and $SO(2,4)$ (with $s = +1$ and $s = -1$, respectively), which is the group that must ultimately be gauged. This extended structure results from broadening the background isometry groups from $SO(1,5)$ to $SO(1,6)$ and from $SO(2,4)$ to $SO(2,5)$. The additional $U(1)$ symmetry is necessitated by the presence of anticommutators and isn't a direct consequence of the background isometry group extensions discussed in Section \ref{sec3.2}.

Given the similar commutation relations for the generators of both algebras in \eqref{backgroundAlgebra} (they only differ by the sign of $s$), we'll focus our subsequent discussion on the gauging of $SO(2,4) \times U(1)$.

We begin by establishing the representation of the algebra generators in the standard four-dimensional Dirac representation. In this framework, the generators of $SO(1,4)$ (and analogously $SO(2,3)$) are constructed from combinations of $\gamma$-matrices, which satisfy the relation:
\begin{equation}
\{\gamma_a,\gamma_b\}=-2\eta_{ab}\mathbbm{1}_4,
\end{equation}
where $a,b=1,\dots,4$, $\eta_{ab}$ is the (mostly plus) Minkowski metric, and $\mathbbm{1}_4$ is the $4 \times 4$ identity matrix.

For the $SO(2,4) \times U(1)$ case, we identify the generators in this representation as follows:
\begin{itemize}
    \item Six Lorentz generators: $M_{ab} = -\frac{i}{4}[\gamma_a,\gamma_b]$
    \item Four translation generators: $P_a = -\frac{1}{2}\gamma_a(1 - \gamma_5)$
    \item Four conformal boost generators: $K_a = \frac{1}{2}\gamma_a(1 + \gamma_5)$
    \item One dilatation generator: $D = -\frac{1}{2}\gamma_5$
    \item One $U(1)$ generator: $\mathbbm{1}_4$
\end{itemize}

These generators fulfill the standard conformal algebra, with the following commutation relations:
\begin{equation}
\label{Lcom}
\begin{aligned}
[M_{ab}, M_{cd}] &= \eta_{bc} M_{ad} + \eta_{ad} M_{bc} - \eta_{ac} M_{bd} - \eta_{bd} M_{ac}, \\
[M_{ab}, P_c] &= \eta_{bc} P_a - \eta_{ac} P_b, \\
[M_{ab}, K_c] &= \eta_{bc} K_a - \eta_{ac} K_b, \\
[P_a, D] &= P_a, \\
[K_a, D] &= -K_a, \\
[K_a, P_b] &= -2(\eta_{ab} D + M_{ab}),
\end{aligned}
\end{equation}
and the following anticommutation relations:
\begin{equation}
\label{Lanticom}
\begin{aligned}
\{M_{ab}, M_{cd}\} &= \frac{1}{2}(\eta_{ac}\eta_{bd} - \eta_{bc} \eta_{ad}) - i \epsilon_{abcd} D, \\
\{M_{ab}, P_c\} &= +i \epsilon_{abcd} P^d, \\
\{M_{ab}, K_c\} &= -i \epsilon_{abcd} K^d, \\
\{M_{ab}, D\} &= 2M_{ab}D, \\
\{P_a, K_b\} &= 4M_{ab}D + \eta_{ab}, \\
\{K_a, K_b\} &= \{P_a, P_b\} = -\eta_{ab}, \\
\{P_a, D\} &= \{K_a, D\} = 0.
\end{aligned}
\end{equation}

\subsection{Fuzzy Gravity}

We've established that four-dimensional noncommutative gravity on $AdS_4$ can be formulated as a gauge theory of the $SO(2,4)\times U(1)$ group\footnote{Noncommutative gravity on $dS_4$ can be handled similarly.}. In order to derive the theory's action, we'll follow the approach outlined in \cite{Manolakos_paper1}. We start by introducing the covariant coordinate $\mathcal{X}_{\mu} \equiv X_{\mu} + A_{\mu}$, where $A_{\mu}$ is the gauge connection, expanded on the gauge group generators as:
\begin{equation}
A_{\mu} = a_{\mu} \otimes \mathbbm{1}_{4} + \omega_{\mu}{}^{ab} \otimes M_{ab} + e_{\mu}{}^{a} \otimes P_{a} + b_{\mu}{}^{a} \otimes K_{a} + \tilde{a}_{\mu} \otimes D.
\end{equation}
Next, we define the appropriate covariant field strength tensor for this theory \cite{Manolakos_paper1, Madore_1992}:
\begin{equation}
\hat{F}_{\mu \nu} \equiv \left[\mathcal{X}_{\mu}, \mathcal{X}_{\nu}\right] - \kappa^2 \hat{\Theta}_{\mu \nu},
\end{equation}
Here, the covariant noncommutativity tensor $\hat{\Theta}_{\mu \nu} \equiv \Theta_{\mu \nu} + \mathcal{B}_{\mu \nu}$ has been introduced, with $\mathcal{B}_{\mu \nu}$ acting as a 2-form field to ensure the correct transformation properties of $\Theta$. Since the field strength is an element of the gauge algebra, it can also be expanded onto the generators of that algebra:
\begin{equation}
\hat{F}_{\mu \nu} = R_{\mu \nu} \otimes \mathbbm{1}_{4} + \frac{1}{2} R_{\mu \nu}{}^{ab} \otimes M_{ab} + \tilde{R}_{\mu \nu}{}^{a} \otimes P_{a} + R_{\mu \nu}{}^{a} \otimes K_{a} + \tilde{R}_{\mu \nu} \otimes D\,.
\end{equation}

To achieve the spontaneous breaking of the $SO(2,4)\times U(1)$ symmetry down to the Lorentz group, we introduce a scalar field $\Phi(X)$. This field transforms in the adjoint (15-dimensional) representation of $SU(4)\sim SO(2,4)$, which is equivalent to transforming as a rank-2 antisymmetric tensor under $SO(2,4)$ \cite{Manolakos_paper1, Manolakos_paper2, Roumelioti:2024lvn}. The scalar must also carry a $U(1)$ charge to ensure that the $U(1)$ symmetry is also broken and doesn't remain in the residual symmetry.

The action is given by:
\begin{equation}
\mathcal{S} = \operatorname{Trtr} \left[\lambda \Phi(X) \epsilon^{\mu \nu \rho \sigma} \hat{F}_{\mu \nu} \hat{F}_{\rho \sigma} + \eta\left(\Phi(X)^2 - \lambda^{-2} \mathbbm{1}_N \otimes \mathbbm{1}_4\right)\right],
\end{equation}
where $\operatorname{Tr}$ is the trace over the matrices representing the coordinates (playing the role of the integration of the commutative case), $\operatorname{tr}$ is the trace over the algebra generators, $\eta$ is a Lagrange multiplier that enforces the constraint on $\Phi$, and $\lambda$ is a parameter with mass dimension. As an element of the gauge group, $\Phi(X)$ can be expanded as:
\begin{equation}
\Phi(X) = \phi(X) \otimes \mathbbm{1}_4 + \phi^{ab}(X) \otimes M_{ab} + \tilde{\phi}^a(X) \otimes P_a + \phi^a(X) \otimes K_a + \tilde{\phi}(X) \otimes D.
\end{equation}
Following the procedure from \cite{Manolakos_paper1, Manolakos_paper2, Roumelioti:2024lvn}, we gauge-fix the scalar field along the dilatation generator:
\begin{equation}
\Phi(X) = \left.\tilde{\phi}(X) \otimes D\right|_{\tilde{\phi} = -2 \lambda^{-1}} = -2 \lambda^{-1} \mathbbm{1}_N \otimes D.
\end{equation}

On-shell, while this condition is satisfied, and after carefully considering the anticommutation relations among the generators and taking traces over their products, the action simplifies significantly to the expression:
\begin{equation}
\mathcal{S}_{br} = \operatorname{Tr}\left(\frac{\sqrt{2}}{4} \varepsilon_{abcd} R_{mn}{}^{ab} R_{rs}{}^{cd} - 4 R_{mn} \tilde{R}_{rs}\right) \varepsilon^{mnrs},
\end{equation}
in which all the additional terms, including the Lagrange multiplier, vanish because of the gauge fixing.

The resulting theory exhibits a residual $SO(1,3)$ gauge symmetry after the spontaneous symmetry breaking. In the commutative limit, where noncommutativity disappears (and with appropriate field redefinitions connecting noncommutative and commutative fields), this action reduces to the Palatini action. This is ultimately equivalent to the Einstein-Hilbert action with a cosmological constant term (as shown in \cite{Manolakos_paper2}). In essence, we recover standard General Relativity with a cosmological constant.

\section{Unification of Conformal and Fuzzy Gravities with Internal Interactions}
\label{sec4}

The minimal way of unifying Conformal Gravity (CG) with internal interactions, specifically those governed by $SO(10)$, is realized by utilizing the $SO(2, 16)$ as the grand unification gauge group. This strategy, as previously noted, draws from the understanding that the dimension of the tangent group doesn't necessarily have to match that of the curved manifold itself \cite{Patellis:2025qbl, roumelioti2407, Roumelioti:2024lvn, Percacci:1984ai, Percacci_1991, Nesti_2008, Nesti_2010, Krasnov:2017epi, Chamseddine2010, Chamseddine2016, noncomtomos, Konitopoulos:2023wst, Weinberg:1984ke}. 

The CG framework naturally arises from gauging $SO(2,4) \sim SU(2,2) \sim SO(6) \sim SU(4)$ (with $SO(6)$ and $SU(4)$ understood in terms of Euclidean signature). Therefore, starting with the $SO(2,16)$ gauge group, one first identifies the centralizer $C_{SO(2,16)}(SO(2,4)) = SO(12)$. Then, this $SO(12)$ is expected to further break down to $SO(10)$, which will serve as the symmetry group for the internal interactions.

For simplicity in the analysis, an Euclidean signature is adopted (the implications of non-compact spaces are discussed in \cite{Roumelioti:2024lvn}). We begin with the group $SO(18)$, placing fermions in its \textbf{256}-dimensional spinor representation. The SSB of $SO(18)$ proceeds initially to its maximal subgroup $SO(6) \times SO(12)$, and subsequently to $SO(6) \times SO(10) \times [U(1)]$, where the brackets on the $U(1)$ are used to take into account both the local and global case. For convenience, let us recall the decomposition of the relevant representations \cite{Slansky:1981yr, Feger_2020, Li:1973mq}:

\begin{equation}\label{so20}
\begin{aligned}
SO(18) & \supset SO(6) \times SO(12) & & \\
{\textbf{18}} & =(\textbf{6},\textbf{1}) + (\textbf{1}, \textbf{12}) & & \text { vector } \\
{\textbf{153}} & =(\textbf{15}, \textbf{1}) + (\textbf{6}, \textbf{12}) + (\textbf{1}, \textbf{66}) & & \text { adjoint } \\
{\textbf{256}} & =({\textbf{4}}, {\overline{\textbf{32}}})+(\overline{\textbf{4}}, {\textbf{32}}) & & \text { spinor } \\
{\textbf{170}} & =({\textbf{1}}, {\textbf{1}})+({\textbf{6}}, {\textbf{12}})+\left({\textbf{20}}^{\prime}, {\textbf{1}}\right)+({\textbf{1}}, {\textbf{77}}) & & \text { 2nd rank symmetric }
\end{aligned}
\end{equation}

The SSB of $SO(18)$ to $SO(6) \times SO(12)$ is accomplished by assigning a vacuum expectation value (VEV) to the $<$$\mathbf{1},\mathbf{1}$$>$ component of a scalar field belonging to the $\mathbf{170}$ representation. Regarding fermions, we begin with the $\mathbf{256}$ spinor representation. 

In order to further break $SO(12)$ down to $SO(10) \times U(1)$ or $SO(10) \times U(1)_{\text{global}}$, we can employ scalar fields from the $\mathbf{66}$ representation (contained within the adjoint $\mathbf{153}$ of $SO(18)$) or the $\mathbf{77}$ representation (contained within the 2nd rank symmetric tensor representation $\mathbf{170}$ of $SO(18)$), respectively, given the following branching rules:
\begin{equation}
\begin{aligned}
SO(12) & \supset SO(10) \times U(1)\\
\textbf{66} & =(\textbf{1})(0)+( \textbf{10})(2)+( \textbf{10})(-2)+( \textbf{45})(0) \\
\textbf{77} & =(\textbf{1})(4)+( \textbf{1})(0)+( \textbf{1})(-4)+( \textbf{10})(2)+( \textbf{10})(-2)+( \textbf{54})(0) \\
\end{aligned}
\end{equation}
Based on the above, a VEV to the $<$$(\mathbf{1})(0)$$>$ component of the $\mathbf{66}$ representation leads to the gauge group $SO(10) \times U(1)$ after SSB. Correspondingly, a VEV to the $<$$(\mathbf{1})(4)$$>$ component of the $\mathbf{77}$ representation results in $SO(10) \times U(1)_{\text{global}}$ after SSB.

Similarly, we can further break $SU(4)$ down to $SO(4) \sim SU(2) \times SU(2)$ in two stages. First, it breaks to $SO(2,3) \sim SO(5)$, and then to $SO(4)$. For this, we recall the following branching rules \cite{Slansky:1981yr}:
\begin{equation}
\begin{aligned}
SU(4) & \supset SO(5)\\
\textbf{4} & =\textbf{4}\\
\textbf{6} & =\textbf{1}+\textbf{5}\\
\end{aligned}
\end{equation}
As an initial step, by assigning a VEV to the $<$$\mathbf{1}$$>$ component of a scalar in the $\mathbf{6}$ representation of $SU(4)$, the latter breaks down to $SO(5)$. Then, according to the branching rules:
\begin{equation}
\begin{aligned}
SO(5) & \supset SU(2) \times SU(2)\\
\textbf{5} & =(\textbf{1},\textbf{1})+(\textbf{2},\textbf{2})\\
\textbf{4} & =(\textbf{2},\textbf{1})+(\textbf{1},\textbf{2})\\
\end{aligned}
\end{equation}
By giving a VEV to the $<$$\mathbf{1},\mathbf{1}$$>$ component of a scalar in the $\mathbf{5}$ representation of $SO(5)$, we ultimately obtain the Lorentz group $SU(2) \times SU(2) \sim SO(4) \sim SO(1,3)$. Additionally, it is notable that in this scenario, the $\mathbf{4}$ representation decomposes under $SU(2) \times SU(2) \sim SO(1,3)$ into the appropriate representations to describe two Weyl spinors.

One can also follow an alternative route to break $SU(4)$ to $SU(2) \times SU(2)$, as discussed in Section \ref{sec2.1}. Specifically, to break the $SU(4)$ gauge group to $SU(2) \times SU(2)$, we can use scalars in the adjoint $\mathbf{15}$ representation of $SU(4)$, which is contained in the adjoint $\mathbf{153}$ representation of $SO(18)$. In this case, we have:
\begin{equation}
\begin{aligned}
SU(4) \supset & SU(2) \times SU(2) \times U(1)\\
\textbf{4} = & (\textbf{2},\textbf{1})(1)+(\textbf{1},\textbf{2})(-1)\\
\textbf{15}  = & (\textbf{1},\textbf{1})(0)+(\textbf{2},\textbf{2})(2)+(\textbf{2},\textbf{2})(-2)\\
&+(\textbf{3},\textbf{1})(0)+(\textbf{1},\textbf{3})(0)\\
\end{aligned}
\end{equation}
Then, by assigning a VEV to the $<$$\mathbf{1},\mathbf{1}$$>$ direction of the adjoint  representation $\mathbf{15}$, we obtain the known result \cite{Li:1973mq} that $SU(4)$ spontaneously breaks to $SU(2) \times SU(2) \times U(1)$. The method for eliminating the corresponding $U(1)$ gauge boson and retaining only $SU(2) \times SU(2)$ was already discussed in Section \ref{sec2.1}. Again, note that the $\mathbf{4}$ representation decomposes into the appropriate representations of $SU(2) \times SU(2) \sim SO(1,3)$ suitable for describing two Weyl spinors.

Having established the analysis of various symmetry breakings using branching rules under maximal subgroups, starting from the group $SO(18)$, one can readily consider instead the isomorphic algebras of the various groups. Specifically, instead of $SO(18)$, one can consider the isomorphic algebra of the non-compact groups $SO(2,16) \sim SO(18)$, and similarly $SO(2,4) \sim SO(6) \sim SU(4)$.

\subsection{Weyl and Majorana conditions on Fermions}

Having explored various SSB patterns in the previous sections, we now turn our attention to fermionic matter fields and examine the implications of imposing Weyl and Majorana conditions in different dimensions and signatures.

A Dirac spinor $\psi$ in $D$ spacetime dimensions has $2^{D/2}$ independent components. The imposition of either the Weyl or the Majorana condition, each reduces this number by a factor of $2$. The Weyl condition is only consistent in even-dimensional spacetimes; therefore, the simultaneous imposition of both conditions, when allowed, results in a Weyl–Majorana spinor with $2^{(D - 2)/2}$ independent components.

The unitary representations of the Lorentz group $SO(1,D-1)$ are labelled by a continuous momentum vector $k$ and a spin `projection' corresponding to a representation of the compact subgroup $SO(D-2)$. The Dirac, Weyl, Majorana, and Weyl–Majorana spinors carry indices that transform as finite-dimensional non-unitary spinor reps of $SO(1,D-1)$.

It is well-known that for non-compact groups $SO(p,q)$, the existence of Majorana–Weyl spinors with signature $(p,q)$ depends on the difference $p-q$. Specifically, such spinors exist if $p - q =0 \mod 8$. To construct a unified theory that includes both CG and internal interactions based on $SO(10)$, and which admits Majorana and Weyl fermions, the minimal group is $SO(1,17)$ \cite{Konitopoulos:2023wst}.

For clarity and to fix notation, let us briefly recall the familiar case of four dimensions. The $SO(1,3)$ spinors in the usual $SU(2) \times SU(2)$ basis transform as $(2, 1)$ and $(1, 2)$, with the representations labelled by their
dimensionality. The two-component Weyl spinors, $\psi_L$ and $\psi_R$, transform as the irreducible spinors, $\psi_L \sim (2, 1)$ and $\psi_R \sim (1, 2)$ of $SU (2) \times SU(2)$, where `$\sim$' here means `transforms as'.

A Dirac spinor is then made by the direct sum of $\psi_L$ and $\psi_R$, $\psi \sim (2,1) \oplus (1,2)$. Accordingly, in four-component notation, the Weyl spinors in the Weyl basis are $(\psi_L, 0)$ and $(0, \psi_R)$ and are eigenfunctions of $\gamma^5$ with eigenvalues $-1$ and $+1$, respectively. 

The usual Majorana condition for a Dirac spinor has the form, $\psi = C \overline{\psi}^T$ , where $C$ is the charge-conjugation matrix. In four dimensions C is off-diagonal in the Weyl basis, since it maps the components transforming as $(\textbf{2}, \textbf{1})$ into $(\textbf{1}, \textbf{2})$. For even D, it is always possible to define a Weyl basis where $\Gamma^{D+1}$ (which consists of the product of all  matrices in D dimensions) is diagonal, therefore
\begin{equation}
\label{gammaEigenV}
    \Gamma^{D+1}\psi_{\pm}=\pm\psi_{\pm}\,.
\end{equation}
We can express $\Gamma^{D+1}$ in terms of the chirality operators in four and extra $d$ dimensions,
\begin{equation}
    \Gamma^{D+1}=\gamma^5\otimes\gamma^{d+1}\,.
\end{equation}
As a result, the eigenvalues of $\gamma^5$ and $\gamma^{d+1}$ are interrelated. It should be noted though, that the choice of the eigenvalue of $\Gamma^{D+1}$ does not impose the eigenvalues on the separate $\gamma^5$ and $\gamma^{d+1}$.

Given that $\Gamma^{D+1}$ commutes with the Lorentz generators, then each of the $\psi_+$ and $\psi_-$ corresponding to its two eigenvalues, according to Eq. \eqref{gammaEigenV}, transforms as an irreducible spinor of $SO(1,D-1)$. For $D$ even, the $SO(1,D-1)$ always has two independent irreducible spinors; for $D = 4n$ there are two self-conjugate spinors $\sigma_D$ and ${\sigma_D}^\prime$, while for $D = 4n + 2$, $\sigma_D$ is non-self-conjugate and $\overline{\sigma}_D$ is the other spinor. Conventionally, it is selected $\psi_-\sim \sigma_D$ and $\psi_+\sim {\sigma_D}^\prime$ or $\overline{\sigma}_D$. Then, Dirac spinors are defined as the direct sum of Weyl spinors,
\begin{equation}
\psi=\psi_+\oplus\psi_-\sim
    \begin{cases}
    \sigma_D\oplus\sigma_D^{\prime} & \mathrm{for}\ D=4n \\
    \sigma_D\oplus\overline{\sigma}_D & \mathrm{for}\ D=4n+2\,.  
    \end{cases}
\end{equation}

The Majorana condition can be imposed in $D=2,3,4+8n$ dimensions and therefore the Majorana and Weyl conditions are compatible only in $D=4n+2$ dimensions. We limit ourselves here in the case that $D = 4n+2$ (or the rest see e.g. refs \cite{CHAPLINE1982461, KAPETANAKIS19924}). Then starting with Weyl–Majorana spinors in $D=4n+2$ dimensions, we are actually forcing a rep, $f_R$, of a gauge group defined in higher dimensions to be the charge conjugate of $f_L$, and we arrive in this way to a four-dimensional theory with the fermions only in the $f_L$ representation of the gauge group.

Let us now consider our case, keeping once more the Euclidean signature. Starting with a Weyl spinor of $\mathrm{SO}(18)$, according to the breakings and branching rules discussed earlier in the present Section we have
\begin{equation}
\begin{aligned}
SO(18) & \supset SU(4) \times SO(12)\\
\textbf{256} & =(\textbf{4},\overline{\textbf{32}})+(\overline{\textbf{4}},\textbf{32}).
\end{aligned}
\end{equation}
Then we have the following branching rule of the \textbf{32} under the $SO(10)\times[U(1)]$:
\begin{equation}
\begin{aligned}
SO(12) & \supset SO(10) \times \left[U(1)\right]\\
\textbf{32} & =\overline{\textbf{16}}\,(1)+\textbf{16}\, (-1).
\end{aligned}
\end{equation}
Recall that the $\left[{U}(1)\right]$ above is there to take into account that the $U(1)$ either remains as a gauge symmetry, or it is broken leaving a $U(1)$ as a residual global symmetry.

On the other hand, as it was noted earlier,
\begin{equation}
\begin{aligned}
SU(4) & \supset SU(2) \times SU(2)\\
\textbf{4} & =(\textbf{2},\textbf{1})+(\textbf{1},\textbf{2}).
\end{aligned}
\end{equation}
Therefore, after all the breakings, we obtain:
\begin{equation}
\begin{gathered}
    SU(2)\times SU(2) \times SO(10) \times [U(1)]\\
    \{(\textbf{2},\textbf{1})+(\textbf{1},\textbf{2})\}\{\textbf{16}(-1)+\overline{\textbf{16}}(1)\}+\{(\textbf{2},\textbf{1})+(\textbf{1},\textbf{2})\}\{\overline{\textbf{16}}(1)+\textbf{16}(-1)\}\\
    =2\times\textbf{16}_L(-1)+2\times\overline{\textbf{16}}_L(1)+2\times\textbf{16}_R(-1)+2\times\overline{\textbf{16}}_R(1), 
\end{gathered}
\end{equation}
from where, given that $\overline{\textbf{16}}_R(1)=\textbf{16}_L(-1)$ and $\overline{\textbf{16}}_L(1)=\textbf{16}_R(-1)$, and finally choosing to keep only the $\textbf{-1}$ eigenvalue of $\gamma^5$, we obtain
\begin{equation}
    4\times \textbf{16}_L(-1)\, .
\end{equation}
Therefore, this construction yields a natural prediction of four fermion families, arising from the underlying group-theoretic structure. The flavour separation is left as an open problem for future work.

Let us recall once more that Weyl spinors can be defined in even dimensions, however Weyl-Majorana spinors can be defined only for $D=2\mod8$ if $F$ is real, and $6 \mod 8$ if $F$ is pseudoreal \cite{CHAPLINE1982461}. It is interesting that Weyl-Majorana spinors can be defined in $SO(1,9)$ and $SO(1,17)$ but not in $SO(2,D-2)$. The latter, if possible, could give a further reduction of the resulting fermions in the present case\footnote{The real and pseudoreal spinors can be obtained from the table 2 of ref \cite{CHAPLINE1982461}.}.

Finally, we briefly comment on the situation in the context of FG. As described in \cite{roumelioti2407}, when pursuing the unification of FG with internal interactions, in analogy to CG unification via $\mathrm{SO}(10)$ \cite{Roumelioti:2024lvn}, we encounter two primary challenges:
\begin{enumerate}
  \item The fermions must remain chiral in order to remain massless at low energies and avoid acquiring Planck-scale masses.
  \item The fermions must belong to matrix (tensorial) representations, since FG is defined as a matrix model.
\end{enumerate}
One can construct a gauge theory with symmetry group $\mathrm{SO}(6) \times \mathrm{SO}(12)$ and fermions in the $(\textbf{4},\overline{\textbf{32}})+(\overline{\textbf{4}},\textbf{32})$ representation, thus satisfying both chirality and matrix representation requirements. When FG is interpreted as a gauge theory of gravity, the gravitational sector naturally aligns with the gauge group $\mathrm{SO}(6) \times \mathrm{U}(1) \sim \mathrm{SO}(2,4) \times \mathrm{U}(1)$. Hence, from this perspective, the deviation from the CG framework is minimal, and a consistent embedding of chiral fermions into the fuzzy model is achieved.

\section{From SO(2,16) to the Standard Model}
\label{sec5}

Four different models that start from the $SO(2,16)\sim SO(18)$ gauge group and lead to the SM are discussed in this section, along with their potential for observation in experiments that look for proton decay and gravitational wave signals.

\subsection{Field content and estimation of symmetry breaking scales}\label{content-scales}

We begin by determining (following \cite{Patellis:2024znm}) the full field content of the  $SO(18)$ gauge theory, from which we  obtain EG and $SO(10)\times [U(1)]_{global}$. We use the breakings and field content of \cite{Djouadi:2022gws}, in order to get the SM from the $SO(10)$ GUT. In particular, $SO(10)$ breaks into an intermediate group, which consequently breaks into the SM group. The intermediate groups are the Pati-Salam, $SU(4)_C\times SU(2)_L\times SU(2)_R$, possibly with a discrete left-right symmetry,~$\mathcal{D}$, and the left-right group, $SU(3)_C\times SU(2)_L\times SU(2)_R\times U(1)_{B-L}$, with or without the $\mathcal{D}$ discrete symmetry. We will denote them 422, 422D, 3221 and 3221D from now on. Thus, for each of the four low-energy cases, we have a distinct field content at the $SO(18)$ level.

According to the previous section, $SO(18)$ breaks into $SO(6)\times SO(12)$ by~the $(\textbf{1},\textbf{1})$ of a scalar $\textbf{170}$ rep and we employ scalars in the $\textbf{15}$ rep of~$SO(6)$ to break CG, which are drawn from the $SO(18)$ rep $\textbf{153}$:
\begin{equation}\label{153}
\textbf{153}   = (\textbf{15},\textbf{1}) + (\textbf{6},\textbf{12}) + (\textbf{1},\textbf{66})~.
\end{equation}

The $SO(12)$ gauge group is spontaneously broken by scalars in the $\textbf{77}$ rep, which can result from the $\textbf{170}$ rep of the $SO(18)$ group. 
In $SO(6)\times SO(12)$ notation, the scalars responsible for breaking this product group belong in $(\textbf{15},\textbf{1})$ and $(\textbf{1},\textbf{77})$. 
Fermions in the $\textbf{16}$ rep of $SO(10)$, are obtained from a $\textbf{256}$ rep of $SO(18)$ (which will~result in the $\textbf{16}$ through $(\overline{\textbf{4}},\textbf{32})$ in $SO(6)\times SO(12)$ notation). The $SO(10)$ GUT is the spontaneously broken by a scalar  in the  $\textbf{210}$ rep into the 422 and the 3221D intermediate groups, by a scalar in the $\textbf{54}$ rep leading to the 422D gauge group and by a scalar in the  $\textbf{45}$ rep  for the 3221 case.
 
Each one of the intermediate gauge groups is broken spontaneously into the SM by scalars in a $\overline{\textbf{126}}$ rep, while the electroweak Higgs boson is accommodated in~a $\textbf{10}$ rep (in $SO(10)$ language). From this point onwards, the scale at which  $SO(10)$ is broken will be called GUT scale, $M_{GUT}$, since all gauge couplings unify at that scale. The scale at which the 422(D)/3221(D) groups break will be referred to as the \textit{intermediate scale}, $M_I$. The consecutive breakings for each case are given:

\footnotesize
\begin{align}
\text{422}: & \quad \text{SO(10)}|_{M_{GUT} } \xrightarrow{\langle \mathbf{210_H} \rangle} \, \, \, SU(4)_C\times SU(2)_R\times SU(2)_R|_{M_{I} }\xrightarrow{\langle \mathbf{\overline{126}_H}\rangle} \text{SM}\, ; \label{breakingchain1} \\
\text{422D}: & \quad \text{SO(10)}|_{M_{GUT} } \xrightarrow{\langle \mathbf{54_H} \rangle} ~~SU(4)_C\times SU(2)_R\times SU(2)_R \times {\cal D}|_{M_{I}}\xrightarrow{\langle \mathbf{\overline{126}_H}\rangle
} \text{SM}  \, ; \label{breakingchain2} \\
\text{3221}: & \quad \text{SO(10)}|_{M_{GUT}} \xrightarrow{\langle \mathbf{45_H} \rangle} \, \, \, SU(3)_C\times SU(2)_L\times SU(2)_R\times U(1)_{B-L}|_{M_{I}}\xrightarrow{\langle \mathbf{\overline{126}_H} \rangle} \text{SM}\, ; \label{breakingchain3}
\\
\text{3221D}: & \quad \text{SO(10)}|_{M_{GUT}} \xrightarrow{\langle \mathbf{210_H} \rangle} SU(3)_C\times SU(2)_L\times SU(2)_R\times U(1)_{B-L}\times {\cal D}|_{M_{I}}\xrightarrow{\langle \mathbf{\overline{126}_H} \rangle} \text{SM} \, .
\label{breakingchain4}
\end{align}
\normalsize
Following the $SO(12)$ branching rules: 
\begin{align} 
     SO(12) &\supset SO(10) \times U(1)_{global} \nonumber\\
   \textbf{12}   &= (\textbf{1})(2) + (\textbf{1})(-2) + ( \textbf{10})(0) \label{12}\\
   \textbf{66}   &= (\textbf{1})(0) + (\textbf{10})(2) +(\textbf{10})(-2) + ( \textbf{45})(0) \label{66}\\
   \textbf{77}   &= (\textbf{1})(4) +(\textbf{1})(0) +(\textbf{1})(-4) + (\textbf{10})(2) +(\textbf{10})(-2) + ( \textbf{54})(0) \label{77}\\
   \textbf{495}   &= (\textbf{45})(0)  + (\textbf{120})(2) +(\textbf{120})(-2) + ( \textbf{210})(0) \label{495}\\
   \textbf{792}   &= (\textbf{120})(0) +(\textbf{126})(0) +(\overline{\textbf{126}})(0) + (\textbf{210})(2) +(\textbf{210})(-2)\label{792}~,
\end{align}
we choose to accommodate~the Higgs $\textbf{10}$ rep into the  $\textbf{12}$ rep of $SO(12)$, while the $\overline{\textbf{126}}$  that breaks the intermediate gauge group comes from $\textbf{792}$. As for the intermediate breakings, $\textbf{45}$  comes from $\textbf{66}$, $\textbf{54}$  from $\textbf{77}$ and $\textbf{210}$  from $\textbf{495}$. 
Considering the $SO(18)$ branching rules:
\begin{align} 
     SO(18) \supset& SO(6) \times SO(12) \nonumber\\
   \textbf{18}   =& ( \textbf{6}, \textbf{1}) + (\textbf{1}, \textbf{12})\label{18}\\
   \textbf{3060}   =& (\textbf{15},\textbf{1}) + (\textbf{10},\textbf{12}) + (\overline{\textbf{10}},\textbf{12}) + (\textbf{15},\textbf{66}) + (\textbf{6},\textbf{220}) + (\textbf{1},\textbf{495})\label{3060}\\
   \textbf{8568}   =& (\textbf{6},\textbf{1}) + (\textbf{15},\textbf{12})+ (\textbf{10},\textbf{66}) + (\overline{\textbf{10}},\textbf{66}) + (\textbf{15},\textbf{220}) + (\textbf{6},\textbf{495}) + \nonumber\\
   &+   (\textbf{1},\textbf{792})\label{8568}~,
\end{align}
 together with (\ref{so20}) and (\ref{153}), the $\textbf{12}$ rep of $SO(12)$ comes from the $\textbf{18}$ rep of $SO(18)$, $\textbf{792}$ from $\textbf{8568}$, $\textbf{66}$~from $\textbf{153}$, $\textbf{495}$ from $\textbf{3060}$ and finally $\textbf{77}$ from $\textbf{170}$.   The full field content with its reps under each gauge group is presented in Table \ref{content}.
%%%%%%%%%%%%%%%%%%%% T A B L E %%%%%%%%%%%%%%%%%%%%%%%%%%%%%%%%%%%%%%%%%%%%%%%
\begin{center}
\begin{table}
\begin{center}\small
\renewcommand{\arraystretch}{2}
\begin{tabular}{|r|r|r|r|}
\hline
    $SO(10)$ & $SO(6)\times SO(12)$ & $SO(18)$  & Type  \& Role  \\\hline
 $\textbf{16}$ & $(\overline{\textbf{4}},\textbf{32})$ & $\textbf{256}$  & fermion   \\\hline
 - & $(\textbf{15},\textbf{1})$ &  $\textbf{153}$ & scalar, breaks $SO(6)$   \\\hline
 - & $(\textbf{1},\textbf{77})$ & $\textbf{170}$ & scalar, breaks $SO(12)$   \\\hline
 $\textbf{10}$ & $(\textbf{1},\textbf{12})$ & $\textbf{18}$ & scalar, breaks SM   \\\hline
 $\overline{\textbf{126}}$ & $(\textbf{1},\textbf{792})$ & $\textbf{8568}$ & scalar, breaks the intermediate  groups into SM  \\\hline
 $\textbf{45}$ & $(\textbf{1},\textbf{66})$ & $\textbf{153}$ & scalar, breaks $SO(10)$ into 3221   \\\hline
 $\textbf{210}$ & $(\textbf{1},\textbf{495})$ & $\textbf{3060}$ & scalar, breaks $SO(10)$ into 422 \& 3221D   \\\hline
 $\textbf{54}$ & $(\textbf{1},\textbf{77})$ & $\textbf{170}$ & scalar, breaks $SO(10)$ into 422D   \\\hline
\end{tabular}\normalsize
\caption{The full field content at each gauge group level. 
}
\label{content}
\end{center}
\end{table}
\end{center}
 %%%%%%%%%%%%%%%%%%%%%%%%%%%%%%%%%%%%%%%%%%
Next, we make an estimation of the scales where each gauge breaking occurs. As mentioned in the previous section, the initial gauge group features an even number of fermionic generations and we chose to have four. This means there will be an extra fermionic generation that is expected to acquire masses and decouple just above the EW scale. However, we can still give a rough estimate of the various breaking scales of our model, based upon the 3gen, 1-loop RG run of \cite{Patellis:2024znm} and assuming that the fourth generation does not significantly qualitatively alter the evolution of the gauge couplings. For all four cases gauge unification is achieved and the intermediate scale is estimated at $10^{10} ~\text{GeV} \lesssim M_I \lesssim 10^{13} ~\text{GeV}$, while the GUT scale at $10^{15} ~\text{GeV} \lesssim M_{GUT} \lesssim 10^{16} ~\text{GeV}$.

The~breaking of the CG (to~EG) gives a negative~contribution to the cosmological constant. Thus, if this was the only contribution,~the space would be AdS. However, there are positive contributions from the $SO(18)$  and $SO(12)$ breakings. By choosing either of these breakings to take place at the same scale with the CG breaking, we can fine-tune the contributions to give a zero or (slightly) positive value of the cosmological constant.

We will focus on three distinct scenarios regarding the breakings above $M_{GUT}$. In scenario \textit{A} the $SO(18)$  group breaks into~$SO(6)\times SO(12)$ and they -in turn- break into EG and $SO(10)$,  all at the same scale,  $M_X$. As such,  the contribution from the breakings of $SO(18)$  cancels the negative to the cosmological constant from the CG breaking. In the scenaria  \textit{B} and \textit{C}, $SO(18)$  breaks~into $SO(6)\times SO(12)$ at a scale $M_B$, while  $SO(6)$ and $SO(12)$ will~break at different scale, $M_X$, between $M_B$ and $M_{GUT}$. In both \textit{B} and \textit{C},  the contribution  from the breaking of $SO(12)$ cancels the negative to the cosmological constant that comes from the CG breaking.

While the RG running couplings below $M_{GUT}$ is straightforward, in the case of gauge theories~based on non-compact~groups the situation is not that clear. Very serious calculations  can be found on  $\beta$-functions regarding Stelle's $R^2$ gravity, which has been proven to~be renormalizable \cite{Stelle:1976gc,Stelle:1977ry}.  However, they are all done in Euclidean space~\cite{Fradkin:1981iu,Avramidi:1985ki,Codello:2006in,Niedermaier:2009zz,Niedermaier:2010zz,Ohta:2013uca}. Thus,~strictly speaking, the $\beta$-functions of a gauge theory~based on a non-compact~group have not been calculated. We can speculate,~though, that, at least~at~one-loop level, these $\beta$-functions can be well~approximated by the respective~ones of their compact counterparts. This is supported by suggestions of Donoghue \cite{Donoghue:2016xnh,Donoghue:2016vck,Donoghue:2017fvm}, which we adapt (see \cite{Patellis:2024znm})).\\

\noindent \textbf{Scenario \textit{A}}:  The $SO(18)$ gauge coupling runs down to the $M_X$ scale, where it should match  both the $SO(6)$ and~$SO(12)$ gauge couplings:
\begin{equation}
\alpha_{10}^{(1)}(M_X)=\alpha_{CG}^{(1)}(M_X)~. \label{mc}
\end{equation}
By substituting the above relation into (\ref{so24finalaction}) and focusing on its last term, we  compare the term with the  contributions to the cosmological~constant that come from $SO(12)$ and get an estimate of the scale:
\begin{equation}
    M_X \sim  10^{18}~\text{GeV}~. \label{MxA}
\end{equation}
However, by running the $SO(18)$ gauge coupling up to $M_{Pl}$, its steep $\beta$-function gets it  in the non-perturbative~regime and it hits a Landau pole before it reaches the Planck scale.\\

\noindent \textbf{Scenaria \textit{B} \& \textit{C}}:  In the first one the $SO(18)$ gauge group breaks below the Planck scale, $M_B<M_{Pl}$, while in the second the $SO(18)$ group is broken at the Planck scale, $M_B=M_{Pl}$. In both scenaria, the $SO(6)\times SO(12)$~gauge group runs down to $M_X$, below which we get $SO(10)$ and EG (plus the global $U(1)$s that we can ignore throughout the study). Again, the $SO(6)$ and $SO(12)$ groups should break at the same scale to fine-tune the cosmological constant.
Using a matching condition like  (\refeq{mc}) is out of the question in either case, since we now have $\alpha_{10}^{(1)}(M_X)=\alpha_{12}^{(1)}(M_X)$. Here, by employing  the $SO(10)$ and $SO(12)$ gauge $\beta$-functions  and the \textit{approximative} gauge $\beta$-function of $SO(6)$, we  make a rough estimate of the scale at $M_X\sim10^{18}$ GeV. 

Above  the $M_X$ scale the $SO(12)$ gauge coupling runs up to $M_B$ and, due to our choice of reps for some of the scalars, it stays within the perturbative regime, as they were chosen in order for the scalars to be always singlets under the CG gauge group and thus to avoid multiplicities in the gauge $SO(12)$ $\beta$-function calculation. In scenario  \textit{B} the $SO(18)$ gauge coupling  should in principle be able to run up to $M_{Pl}$ while staying in the perturbative regime.

We close this subsection with recalling our previous comment on the FG case, as it was discussed in Section \ref{sec4}.

%The attempt to unify FG with internal interactions along the lines of \cite{Roumelioti:2024jib} and \cite{Roumelioti:2024lvn} finds several difficulties. Fermions should be chiral to avoid Planck scale masses and, since FG is a matrix model, they should appear in a matrix rep. The Majorana condition can be imposed and as such a solution could be to start with the $SO(6) \times SO(12)$ group as the initial gauge theory with the fermions  accommodated in the $(\overline{\textbf{4}},\textbf{32})$ rep. Following the gauge-theoretic formulation of gravity to construct a FG, we gauge $SO(6) \times U(1) \sim SO(2,4) \times U(1)$. This makes it very similar to the  CG case  and we may identify the FG model to scenario \textit{C}, with the difference that there is no $SO(18)$ above $M_{Pl}$.

\subsection{Cosmic Strings from intermediate scale symmetry breakings and constraints from proton decay}\label{cosmic-strings}

Although proton decay has not yet been observed, the proton lifetime has been the focus of many experimental searches \cite{Super-Kamiokande:2013rwg, Super-Kamiokande:2014otb, Super-Kamiokande:2016exg, Heeck:2019kgr} which constrain it and, consequently, the GUT scale. Most of the breaking paths of  $SO(10)$ are tested by Super-Kamiokande (Super-K), while future experiments like Hyper-Kamiokande (Hyper-K) \cite{Hyper-Kamiokande:2018ofw}, DUNE \cite{DUNE:2016hlj} and JUNO \cite{JUNO:2015zny} will improve the sensitivity by even one order~of magnitude. These experiments could be getting very~close to the proton decay observation and, in turn, baryon  number violation, a development that could in principle exclude many attempts of grand unification.

However, there are current and future experiments focusing also on other paths than proton decay to probe GUTs.

When a gauge structure breaks spontaneously down to the SM, it produces topological defects. Out of the three most prominent topological defects, domain walls and monopoles  dominate the energy density of the universe and are thus problematic, although this can be ameliorated by the assumption that inflation happens after their production, as it strongly suppresses their density. Cosmic strings, on the other hand, are usually formed from the breaking of an abelian $U(1)$~subgroup and do not feature such problems, since a cosmic string~network has a~scaling solution and consequently does~not overclose~the~Universe, but it can survive and generate~a source~of~gravitational radiation~\cite{Vilenkin:1984ib, Caldwell:1991jj, Hindmarsh:1994re}. 

Gravitational waves (GWs) originating from cosmic strings have been recently pointed out as a way to probe high-energy  models  \cite{Damour:2001bk, Damour:2004kw, Dror:2019syi, Buchmuller:2019gfy, Chakrabortty:2020otp, King:2020hyd}. If inflation takes place before the formation of cosmic strings, when they~intersect and form~loops their network becomes a GW source. When they transition between different states,~they emit~strong high-frequency GW beams and the loops emit~energy in a gravitational manner. This radiation is called stochastic gravitational~wave background (SGWB). An approach to compute SGWB can be found in \cite{Cui:2018rwi}

In our study  we demonstrate four breaking chains that lead from the $SO(10)$  GUT to the SM, denoted 422, 422D, 3221 and 3221D respectively, and we estimate the intermediate scale $M_I$ and the unification scale $M_{GUT}$. \cite{King:2021gmj} calculates in 2-loop level the intermediate and unification scales for all the breaking chains of $SO(10)$, including the four cases at hand. Although our examples feature a fourth fermionic generation, we follow their analysis (which compares their~numerical results to the experimental bounds on proton decay/lifetime of Super-K and Hyper-K) in an approximative manner, as above. Neither 422D  nor 3221D satisfy the Super-K bounds, while 422 and 3221 are just above the lower bounds for $M_{GUT}$. However, during the future Hyper-K run proton decay~is not~observed, 3221 would be excluded as well. As such, 422 is the candidate with the highest survivability.

Turning our attention to the production of topological defects from spontaneous symmetry breakings (see \cite{King:2021gmj} for details), the $SO(10)$ gauge group breaking leads to~monopoles in all cases and additional cosmic strings~in 422D and 3221D. Therefore inflation should happen after this breaking and, since it washes them out,  their gravitational signal is rendered undetectable. Considering the intermediate breakings, we have production of monopoles from the 422 gauge group, monopoles and domain walls from 422D, cosmic strings from 3221 and~cosmic strings and domain walls from 3221D. Following the above reasoning, inflation has to happen after the~breakings of 422, 422D~and 3221D and thus these cases cannot be probed  through GW background. However,~in the 3221 case, inflation can strategically take place between the GUT and 3221 breakings and therefore strings can in principle be~observed through SGWB. The tension of cosmic strings that is generated from the 3221 breaking is estimated to be compatible with the Super-K bounds. Should proton decay not be detected in Hyper-K, this channel should also be excluded.

\section{Conclusions}

We have presented a comprehensive framework for a possible unification of gravity(ies) with internal interactions. This is primarily based on the recent suggestion of~\cite{Roumelioti:2024lvn}, namely that such a unification can be achieved by gauging an enlarged tangent Lorentz group. This idea is motivated by the observation that the dimension of the tangent group need not coincide with that of the curved manifold. Moreover, since gravity can be described by gauge theories—similarly to the SM—a natural proposal emerges: both gravitational and internal interactions could originate from a common gauge structure, as captured in the four-dimensional unification scheme developed here.

The gravity theories considered in the present work include CG and FG, both based on gauging the conformal group $SO(2,4)$ (FG also requires gauging an additional $U(1)$). A significant result is that CG can be spontaneously broken either to EG or to WG, with WG also capable of eventually breaking to EG.

The unification of CG and FG with internal interactions in four dimensions was then constructed by gauging the higher-dimensional tangent group $SO(2,16)$, which is successively broken to EG and the $SO(10)$ GUT. Fermion inclusion, along with suitable application of the Weyl condition, results in a fully unified setting. A 1-loop analysis at low energies, exploring four possible breaking channels of $SO(10)$ down to the SM, yields estimates of all relevant breaking scales, ranging from the Planck to the EW scale.

Clearly both CG and FG describe physics close to Planck scale, as it was estimated in Section \ref{sec5}, while the GUT $SO(10)$ and the Standard Model of Particle Physics appears in lower energies. In lower energies (see estimates in Section \ref{sec5}) the CG is broken to EG, while FG first goes to its continuum limit and then behaves very similarly to CG. In both cases the low energy regime in which the GUT $SO(10)$ and the SM are discussed is the usual framework of EG on the gravity side.

Finally, we estimated the experimental prospects of each breaking channel, focusing on gravitational wave signals and proton decay, following previous analyses. Two channels are excluded by proton lifetime bounds. Among the remaining two, one predicts the formation of cosmic strings which generate a stochastic gravitational wave background with detectable signal. Thus, the proposed framework is not only theoretically robust but also, in principle, potentially testable in experiments.

Although the present unified scheme belongs to the category that Stelle \cite{Stelle:1976gc} has proven to be renormalizable, we do not claim yet that it is renormalizable. Similarly, we do not claim that is ghost free, although we have been involved in \cite{Hell:2023rbf} with an examination of the conditions for removing the ghost. We plan to return to these serious theoretical questions, as well as to further examination of cosmological consequences of the presented unified scheme in future publications.

\section*{Acknowledgements}
It is a pleasure to thank Costas Bachas, Thanassis Chatzistavrakidis, Jean-Pierre Derendinger, José Figueroa-O'Farrill, Alex Kehagias, Tom Kephart, Spyros Konitopoulos, Dieter Lust, George Manolakos, Pantelis Manousselis, Carmelo Martin, Tomás Ortín, Roberto Percacci, Manos Saridakis and Nicholas Tracas, for our discussions on various stages of development of the theories presented in the current work.

DR would like to thank NTUA for a fellowship for doctoral studies. GZ would like to thank the Arnold Sommerfeld Centre - LMU Munich for their hospitality and support, the University of Hamburg and DESY for their hospitality, and the CLUSTER of Excellence ``Quantum Universe'' for support. GP would like to thank the Institute of Physics of U.N.A.Mexico for their warm hospitality.

\section*{Author Contributions}
Conceptualization, George Zoupanos; Software, Gregory Patellis; Writing – original draft, Gregory Patellis, Danai Roumelioti and Stelios Stefas; Writing – review and editing, Gregory Patellis, Stelios Stefas and George Zoupanos; Supervision, George Zoupanos. All authors have read and agreed to the published version of the manuscript.

\section*{Funding}
This research received no external funding.

\section*{Data Availability Statement}
The original contributions presented in
this study are included in the article/supplementary material. Further
inquiries can be directed to the corresponding author.

\section*{Conflicts of Interest}
The authors declare no conflicts of interest.

\appendix
\section{Brief Historical review}
\label{appA}
This section provides a concise historical overview of concepts pioneered by Snyder \cite{Snyder:1946qz} and Yang \cite{yang1947}. This overview helps to illuminate the rationale behind the construction of the background space detailed in Section \ref{sec3.2}.

We begin by examining Snyder's contribution \cite{Snyder:1946qz}. Snyder was the first to propose a quantized, or discretized, model of spacetime that maintained Lorentz symmetry. He achieved this by introducing a fundamental length scale and equating spacetime coordinates with elements of the Lie algebra of the de Sitter group.

Specifically, Snyder considered the four-dimensional de Sitter group $SO(1,4)$, whose generators adhere to the following Lie algebra:
\begin{equation}
\left[J_{\mu \nu},J_{\rho \sigma}\right]=i\left(\eta_{\mu \rho}J_{\nu \sigma}+\eta_{\nu \sigma}J_{\mu \rho}-\eta_{\nu \rho}J_{\mu \sigma}-\eta_{\mu \sigma}J_{\nu \rho}\right)\, ,
\end{equation}
where $\mu,\nu,\rho,\sigma = 0, \dots, 4$, with $J_{\mu \nu} = -J_{\nu \mu}$ and $\eta_{\mu \nu}$ representing the five-dimensional Minkowski metric with a signature of $\operatorname{diag}(-,+,+,+,+)$.

By decomposing $SO(1,4)$ into its largest subgroup, $SO(1,3)$, the algebra yields three distinct relations:
\begin{equation}
\begin{gathered}
\left[J_{ij},J_{kl}\right]=i\left(\eta_{i k}J_{j l}+\eta_{j l}J_{i k}-\eta_{j k}J_{i l}-\eta_{i l}J_{j k}\right),\\ \left[J_{i j},J_{k4}\right]=i\left(\eta_{i k}J_{j4}-\eta_{j k}J_{i4}\right),\ \left[J_{i4},J_{j4}\right]=i J_{ij},\
\end{gathered}
\end{equation}
with indices $i,j,k,l = 0, \dots, 3$. To connect these generators to physical observables, Snyder made the following identifications:
\begin{equation}
\label{identifications.X,TH}
\Theta_{ij} = \hbar J_{ij}, \quad X_i = \lambda J_{i4},
\end{equation}
where $\lambda$ signifies a fundamental length scale. This leads to the commutation relations:
\begin{equation}
\begin{gathered} \left[\Theta_{ij},\Theta_{kl}\right] = i\hbar\left(\eta_{i k}\Theta_{j l}+\eta_{j l}\Theta_{i k}-\eta_{j k}\Theta_{i l}-\eta_{i l}\Theta_{j k}\right),\\ \left[\Theta_{ij},X_k\right] = i\hbar\left(\eta_{i k}X_j - \eta_{j k}X_i\right),\ \left[X_i,X_j\right] = \frac{i\lambda^2}{\hbar}\Theta_{ij},\
\end{gathered}
\end{equation}
which explicitly illustrates the noncommutative nature of the spacetime coordinates.
\vspace{45pt}

Building on Snyder’s foundational work, Yang investigated the possibility of incorporating continuous translations into a noncommutative spacetime model \cite{yang1947}. His approach expanded the algebraic structure by considering the higher symmetry group $SO(1,5)$ \cite{Snyder:1946qz, kastrup_1966, Heckman_2015}, whose generators satisfy:
\begin{equation} \left[J_{mn},J_{rs}\right]=i\left(\eta_{mr}J_{ns}+\eta_{ns}J_{mr}-\eta_{nr}J_{ms}-\eta_{ms}J_{nr}\right),
\end{equation}
where $m,n,r,s = 0, \dots, 5$, and $\eta_{mn} = \operatorname{diag}(-1,1,1,1,1,1)$.
By sequentially decomposing $SO(1,5)$ down to $SO(1,3)$—specifically, through the chain $SO(1,5) \supset SO(1,4) \supset SO(1,3)$—the algebra yields nine distinct relations:
\begin{equation}
\begin{gathered}
\left[J_{ij},J_{kl}\right]=i\left(\eta_{i k}J_{j l}+\eta_{j l}J_{i k}-\eta_{j k}J_{i l}-\eta_{i l}J_{j k}\right),\ \left[J_{ij},J_{k5}\right]=i\left(\eta_{i k}J_{j5}-\eta_{j k}J_{i5}\right),\\ \left[J_{i5},J_{j5}\right]=i J_{ij},\ \left[J_{ij},J_{k4}\right]=i\left(\eta_{i k}J_{j4}-\eta_{j k}J_{i4}\right),\ \left[J_{i4},J_{j4}\right]=i J_{ij},\ \left[J_{i4},J_{j5}\right]=i \eta_{ij}J_{45},\\ \left[J_{ij},J_{45}\right]=0,\ \left[J_{i4},J_{45}\right]=-i J_{i5},\ \left[J_{i5},J_{45}\right]=i J_{i4},\
\end{gathered}
\end{equation}
As before, we establish a connection between these generators and physical quantities through the identifications:
\begin{equation} \Theta_{ij} = \hbar J_{ij}, \quad X_i = \lambda J_{i5},
\end{equation}
and define momenta as:
\begin{equation} \label{identifications.P} P_i = \frac{\hbar}{\lambda} J_{i4}, \end{equation}
while also setting $h=J_{45}$. These identifications lead to the following commutation relations:
\begin{equation}
\begin{gathered}
\left[\Theta_{ij},\Theta_{kl}\right] = i \hbar\left(\eta_{i k}\Theta_{j l}+\eta_{j l}\Theta_{i k}-\eta_{j k}\Theta_{i l}-\eta_{i l}\Theta_{j k}\right),\ \left[\Theta_{ij},P_k\right] = i\hbar\left(\eta_{i k}P_j - \eta_{j k}P_i\right),\\ \left[P_i,P_j\right] = \frac{i \hbar}{\lambda^2} \Theta_{ij},\ \left[\Theta_{ij},X_k\right] = i\hbar\left(\eta_{i k}X_j - \eta_{j k}X_i\right),\ \left[X_i,X_j\right] = \frac{i\lambda^2}{\hbar} \Theta_{ij},\\ \left[X_i,P_j\right] = i\hbar \eta_{ij} h,\ \left[\Theta_{ij},h\right] = 0,\ \left[X_i,h\right] = \frac{i\lambda^2}{\hbar} P_i,\ \left[P_i,h\right] = -\frac{i\hbar}{\lambda^2} X_i.\
\end{gathered}
\end{equation}

From these relations, two vital implications emerge. First, with momenta now integrated into the Lie algebra, they too become noncommutative, signaling a quantization of momentum space. Second, the commutation relations between positions and momenta naturally produce a Heisenberg-like uncertainty structure, aligning with the predictions of quantum mechanics.

\printbibliography

\end{document}